\documentclass[aps,prd,twocolumn,showpacs,superscriptaddress,groupedaddress,nofootinbib]{revtex4}

\usepackage{fancyhdr}
\usepackage{amsthm}
\usepackage{amsmath,amsthm,amssymb}
\usepackage{enumerate}
\usepackage{color}
\usepackage{subcaption}
\usepackage{color}
\usepackage{relsize}
\usepackage{graphicx}
\usepackage{caption}
\usepackage{subcaption}

\makeatletter
\def\@fpheader{\relax}
\makeatother

\def\beq{\begin{equation}}
\def\eeq{\end{equation}}
\def\bea{\begin{eqnarray}}
\def\eea{\end{eqnarray}}

\newcommand{\vecx}{\mathbf{x}}
\newcommand{\vecy}{\mathbf{y}}
\newcommand{\vecp}{\mathbf{p}}
\newcommand*\xbar[1]{%
  \hbox{%
    \vbox{%
      \hrule height 0.5pt % The actual bar
      \kern0.5ex%         % Distance between bar and symbol
      \hbox{%
        \kern 0 em%      % Shortening on the left side
        \ensuremath{#1}%
        \kern-0.1em%      % Shortening on the right side
      }%
    }%
  }%
}

\begin{document}

\title{Dark Matter From Spacetime Nonlocality}

\author{Mehdi Saravani}\email{msaravani@pitp.ca}
\affiliation{Perimeter Institute
for Theoretical Physics, 31 Caroline St. N., Waterloo, ON, N2L 2Y5, Canada}
\affiliation{Department of Physics and Astronomy, University of Waterloo, Waterloo, ON, N2L 3G1, Canada}
\author{Siavash Aslanbeigi}\email{saslanbeigi@pitp.ca}
\affiliation{Perimeter Institute
for Theoretical Physics, 31 Caroline St. N., Waterloo, ON, N2L 2Y5, Canada}

%\date{\today}
%%%%%%%%%%%%%%%%%%%%%%%%%%%%%%%%%%%%%%%
%%%%%%%%%%%%%%%%%%%%%%%%%%%%%%%%%%%%%%%
%%%%%%%%%%%%%%Abstract%%%%%%%%%%%%%%%%%%%%%
%%%%%%%%%%%%%%%%%%%%%%%%%%%%%%%%%%%%%%%
%%%%%%%%%%%%%%%%%%%%%%%%%%%%%%%%%%%%%%%

\begin{abstract}
We propose that dark matter is not yet another new particle in nature, 
but that it is a remnant of quantum gravitational effects on known fields.
We arrive at this possibility in an indirect and surprising manner: 
by considering retarded, nonlocal, and Lorentzian
evolution for quantum fields. This is inspired by recent developments
in causal set theory, where such an evolution shows up
as the continuum limit of scalar field propagation on a background causal set. 
Concretely, we study the quantum theory of a massless scalar field
whose evolution is given not by the
the d'Alembertian $\Box$, but by an operator $\widetilde \Box$ which is Lorentz invariant,
reduces to $\Box$ at low energies, and defines an explicitly retarded evolution:
$(\widetilde \Box \phi)(x)$ only depends on $\phi(y)$, with $y$ is in the causal past of $x$.
This modification results in the existence of a continuum of massive particles,
in addition to the usual massless ones, in the free theory. 
When interactions are introduced,  
these massive or off-shell quanta can be produced by the scattering of massless particles,
but once produced, they no longer interact, which makes them a natural 
candidate for dark matter.

%In this paper, we examine the physical consequences of non-local, causal, and Lorentzian propagation of a massless scalar field.
%%Local propagation is described by $\Box=\eta^{\mu \nu}\partial_{\mu}\partial_{\nu}$\footnote{most pluses signature} which has been replaced by a general non-local operator $\widetilde \Box$ in this paper. Although we give up on local propagation, we still require operator $\widetilde \Box$ to be Poincare-invariant. 
%This has been done by replacing d'Alembertian operator, $\Box=\eta^{\mu \nu}\partial_{\mu}\partial_{\nu}$, with a general non-local operator $\widetilde \Box$, such that $(\widetilde \Box \phi)(x)$ only depends on the value of $\phi$ in the causal past of point $x$.
%We impose minimal requirements on $\widetilde \Box$ to recover local propagation at low energies, and ensure the stability of evolution. 
%Then, a quantum field theory based on $\widetilde \Box$ is proposed.
%%We develop a quantum theory based on such an operator has been discussed. 
%We show that this modification results into the existence of a continuum of massive particles in addition to massless particles in the free theory. Moving to the interacting theory, we demonstrate that the cross section of any scattering with massive modes is zero. As a result, although these new modes can be produced through scattering, they do not take part in any. Consequently, there is no way of detecting them through scattering experiments. This behaviour makes them a perfect candidate to be considered as dark matter.
\end{abstract}
\maketitle
\flushbottom

%==========================================================
\section{Introduction}
%==========================================================
The nature of dark matter is one of the most important problems in modern physics. 
Almost a century after it was hypothesized, though, our understanding of it is still limited to 
its gravitational signature on luminous matter. 
It is often assumed that dark matter is a new weakly interacting particle which is just hard to detect. 
However, so far there has been no conclusive direct or indirect detection in accelerators or 
cosmological/astrophysical settings. 
In what follows, we propose that dark matter is not yet another new particle in nature, 
but that it is a remnant of quantum gravitational effects on known fields.
We arrive at this possibility in an indirect and surprising manner: 
by considering retarded, nonlocal, and Lorentzian
evolution for quantum fields. Concretely, we study the consequences of replacing
the d'Alembertian $\Box$ with an operator $\widetilde \Box$ which is Lorentz invariant,
reduces to $\Box$ at low energies, and defines a retarded evolution:
$(\widetilde \Box \phi)(x)$ only depends on $\phi(y)$, with $y$ is in the causal past of $x$.
Why is this type of evolution interesting, what does it have to do with quantum gravity,
and how does it lead to a proposal for the nature of dark matter?

The causal set theory approach to quantum gravity 
postulates that the fundamental structure of spacetime 
is that of a locally finite and partially ordered set \cite{Sorkin_1}.
Its marriage of discreteness with causal order implies that
physics cannot remain local at all scales.
This nonlocality manifests itself concretely, for instance, 
when one seeks to describe the wave propagation of a scalar 
field on a causal set.
It has been shown in this case that coarse-graining the quantum gravitational 
degrees of freedom leads to a nonlocal field theory described
by an operator exactly of the type $\widetilde \Box$ \cite{2007gr.qc.....3099S, Aslanbeigi:2014zva, 2010PhRvL.104r1301B,
2013CQGra..30s5016D, 2014CQGra..31i5007G}. 
There are reasons to suspect that 
this type of nonlocality
is not necessarily confined to the Planck scale, and that it may have nontrivial
implications for physics
at energy scales accessible by current experiments (see \cite{Woodard:2014iga, Deser:2007jk} and references therein for implications of nonlocality in the context of cosmology).
It is then only natural to wonder what a quantum field theory built upon
$\widetilde \Box$ would look like, especially that it may contain information
about the fundamental structure of spacetime.

Studying $\widetilde \Box$ is also interesting from a purely field-theoretic perspective,
since it forces us to relax one of the core assumptions of 
quantum field theory: locality. 
Most nonlocal and Lorentzian quantum field theories
studied in the literature consider modifications of the type $\Box\to f(\Box)$.
In this paper, we consider explicitly retarded operators, which are more generic
and have more interesting properties as a result. For instance, the Fourier
transform of $\widetilde \Box$ is generically complex, which is a direct consequence of
retarded evolution. In fact, this feature is at the heart of our proposal for 
the nature of dark matter. It is also worth mentioning that
quantizing a field theory of the type described here is non-trivial due to the
absence of a local action principle. This presents a technical challenge,
from which one may gain deeper insight into quantization 
schemes.

What is the relation between a quantum field theory based on $\widetilde \Box$ and 
dark matter? Upon quantizing a free massless scalar field $\phi(x)$ with the classical
equation of motion $\widetilde \Box\phi(x)=0$, 
we find {\it off-shell modes} in the 
mode expansion of the quantized field operator $\widehat{\phi}(x)$.
These are modes which do not satisfy any dispersion relation,
unlike in usual local quantum field theory (LQFT) where every Fourier mode
with four-momentum $p$ is an on-shell quanta, i.e. it satisfies $p\cdot p=0$.%
\footnote{
		We use a signature of $-+++$ for the Minkowski metric $\eta_{\mu\nu}$. Also,
		$p_1\cdot p_2\equiv\eta_{\mu\nu}p_1^{\mu}p_2^{\nu}.$
	     }
This is equivalent to the statement that the quantized field operator
does not generically satisfy the classical equation of motion: $\widetilde \Box\widehat{\phi}(x)\neq0$.
Note that an off-shell mode of a massless scalar field
has an effective mass, and can be thought of as a massive quanta in itself.
We show that the off-shell modes can exist in ``in" and ``out" states of scattering,
and are different from virtual particles which exist as intermediate states in 
Feynman diagrams. 
When considering the interacting theory, we find an extremely surprising result:
the cross-section of any scattering process which contains 
one or more off-shell particle(s)\footnote{In the quantum theory, an off-shell particle is  1-particle quantum state with a well-defined (non-zero) mass and momentum, i.e. a massive eigenstate of Hamiltonian and momentum operator.} in the ``in" state is zero. That is to say, {\it on-shell quanta
can scatter and produce off-shell particles,
but once produced, off-shell particles no longer interact}. It is this behaviour that makes
these off-shell particles a natural candidate for dark matter.
The phenomenological story would be that dark matter particles were produced in the early
universe in this fashion: as off-shell modes of quantum fields. 
This feature of the theory can be traced back to the fact that $\widetilde \Box$ defines an explicitly retarded
evolution, which as mentioned previously, may be a remnant of quantum gravitational 
degrees of freedom.

Our paper is organized as follows. In Section \ref{mBoxDef},
we start by setting forth a series of axioms which any non-local, retarded, and
Lorentzian modification of $\Box$ at high energies should satisfy.
In Section \ref{classicalTheory}, we argue there is no action principle for the
theory of interest, which forces us to carefully study, in Section \ref{QuantumFormalism},
what quantization scheme should be used. There, we argue that canonical quantization
and the Feynman path-integral approach do not work, and explain why the Schwinger-Keldysh 
(also known as the double path integral or in-in) formalism provides the appropriate
framework.
Sections \ref{interaction} and \ref{cross-section} describe the interacting theory,
where we work out the modified Feynman rules, find S-matrix amplitudes, and compute cross-sections
for various examples and comment on the time reversibility of the theory. Although a {\it continuum superposition} of off-shell particles can in principle scatter into on-shell modes, we argue why this is unlikely to happen. 
Extension to massive scalar fields is discussed in \ref{massive}. Section \ref{conclusion} concludes the paper.
\section{Modified d'Alembertian: Definition}
\label{mBoxDef}
%==========================================================
In this section we study generic spectral properties of non-local and Lorentzian 
modifications of the d'Alembertian $\Box$. We focus on a class of 
operators $\widetilde \Box$ which defines an explicitly retarded evolution: 
$(\widetilde \Box\phi)(x)$ depends only on $\phi(y)$ with $y$ in the causal past of $x$. 
As we will see, such operators have interesting features which are absent in
modifications of the type $f(\Box)$.
We start by setting forth a series of axioms which a non-local, retarded, and
Lorentzian modification of $\Box$ at high energies should satisfy:
%will consider all operators $\widetilde \Box$ which satisfy the following axioms:
%We require $\widetilde \Box$ to satisfy the following conditions:
\begin{enumerate}
\item 
{\textbf{Linearity}:
	\beq\label{linearity}
		\widetilde \Box(a \phi+b\psi)=a\widetilde \Box\phi+b\widetilde \Box\psi,\qquad a,b\in \mathbb{C}, 
	\eeq
	where $\phi$ and $\psi$ are complex scalar fields and 
	$\mathbb{C}$ denotes the set of complex numbers.
}

\item
{\textbf{Reality}: 
	for any real scalar field $\phi$, $\widetilde \Box\phi$ is also real. 
	Note that reality and linearity imply for any complex scalar field $\phi$ that
	\beq\label{complexconj}
	(\widetilde \Box\phi^*)=(\widetilde \Box\phi)^*,
	\eeq
 	where $*$ denotes complex conjugation.
}
	
\item 
{\textbf{Poincare-invariance}: 
	evolution defined by $\widetilde \Box$ is Poincare-invariant.
	 Consider a scalar field $\phi(x)$ which transforms to $\phi'(x)=\phi(\Lambda^{-1}x)$ 
	under a Poincare transformation $x\to\Lambda x$.
	We require $\widetilde \Box$ to be invariant under the action of $\Lambda$:
	% means that the action of $\widetilde \Box$ 	
	%on $\phi'(x)$ should be the same as performing $\Lambda$ transformation on $(\widetilde \Box\phi)(x)$. 
	%In other words
	\beq\label{pinv}
		(\widetilde \Box\phi')(x)=(\widetilde \Box\phi)(\Lambda^{-1}x).
	\eeq
	Taking $\Lambda$ to be a spacetime translation $\Lambda(x)=x+a$, one finds that
	the eigenfunctions of $\widetilde \Box$ are plane waves. To see this, let $\phi(x)=e^{ip\cdot x}$
	and define $\psi(x)\equiv(\widetilde \Box\phi)(x)$. It then follows from \eqref{pinv} that
	\beq
		e^{-ip\cdot a}\psi(x)=\psi(x-a),
	\eeq
	where we have used the linearity condition. Solutions to the above equation are 
	plane waves:
	\beq\label{translationReq}
		\psi(x)=\widetilde \Box e^{ip\cdot x}=B(p)e^{ip\cdot x},
	\eeq
	where $B(p)$ is any function of the wave-vector $p$. Therefore, 
	it follows from translational invariance that $e^{ip\cdot x}$ is an 
	eigenfunction of $\widetilde \Box$ with the corresponding eigenvalue $B(p)$ .
	Taking $\Lambda$ to be a Lorentz transformation, it can be shown that 
	$B(p)$ can only depend on the 
	the Lorentzian norm of $p$, i.e. $p\cdot p\equiv\eta_{\mu\nu}p^{\mu}p^{\nu}$, and  
	whether or not $p$ is future or past directed, i.e. $\text{sgn}(p^0)$:
	\beq\label{LIreq}
		B(p)=B(\text{sgn}(p^0),p\cdot p).
	\eeq	
	Combining \eqref{translationReq} and \eqref{complexconj} we find $B(-p)=B^*(p)$,
	which using \eqref{LIreq} is equivalent to
	\beq\label{eq3}
		B(-\text{sgn}(p^0),p\cdot p)=B(\text{sgn}(p^0),p\cdot p)^*.
	\eeq
	For a spacelike wave-vector $p^{\mu}$, it is always possible to find a coordinate system in which 
	$p^0=0$. As a result, $B(p)$ is real for spacelike $p$. For timelike momenta, however, 
	%$\widetilde \Box$ generically 	differentiates between future and past directed plane-waves: 
	\textit{$B(p)$ may be complex and its imaginary part changes sign when $p^0\rightarrow -p^0$}.

	Most nonlocal modifications of $\Box$ considered in the literature are of the form $f(\Box)$,
	in which case $B(p)$ is only a function of $p\cdot p$. In this paper we focus on
	a class of nonlocal operators for which $B(p)$ does depend on $\text{sgn}(p^0)$,
	and find many interesting consequences as a result.
}

\item %Locality at low energies: 
{\textbf{Locality at low energies}:
%We want to recover local evolution at low energies. It means at low energies $\widetilde \Box \approx \Box$. 
	since $\Box$ provides a good description of nature at low energies, we require
	$\widetilde \Box \to \Box$ in this regime.
	In other words, expanding $B(\text{sgn}(p^0),p\cdot p)$ 
	%in terms of $p\cdot p$ 
	for ``small" values of $p \cdot p$, 
	we require the leading order behaviour to be that of $\Box$: %to be $-p \cdot p$:
	\beq\label{IRcondition}
		B(p)\xrightarrow{p\cdot p\rightarrow 0}-p \cdot p.
	\eeq
	Note that by a ``small" value of $p\cdot p$, we mean in comparison to a scale which can be 
	interpreted as the non-locality scale, implicitly defined through $\widetilde \Box$.
}
\item 
{\textbf{Stability}: 
	we require that evolution defined by $\widetilde \Box$ is stable. 
	This condition implies that $B(p)$, when analytically continued to the complex plane of $p$, 
	only has a zero at $p\cdot p=0$ \cite{Aslanbeigi:2014zva}.
}

\item
{\textbf{Retardedness}:
$(\widetilde \Box \phi)(x)$ only depends on $\phi(y)$, with $y$ is in the causal past of $x$.

}
\end{enumerate}
%\begin{figure}
%\centering
%\includegraphics[width=0.9\textwidth]{greenIntg.pdf}
%\caption{This figure shows the analytic structure of $B(p)$ in the complex plane of $p^0$. Generically, there is a cut for time-like momenta. The value of $B(p)$ for time-like momenta corresponds to the value above the cut.}
%\label{analytic}
%\end{figure}

Let us briefly consider a class of operators which satisfy all the aforementioned axioms. 
We shall let $\Lambda$ denote the nonlocality energy scale and define
\beq
\Lambda^{-2}(\widetilde \Box\phi)(x)=a\phi(x)+\Lambda^4\int_{J^{-}(x)}f(\Lambda^2\tau_{xy}^2)\phi(y)d^4y,
\label{mBoxEX}
\eeq
where $a$ is a dimensionless real number, $J^-(x)$ denotes the causal past of $x$,
and $\tau_{xy}$ is the Lorentzian distance between $x$ and $y$: 
\beq
\tau_{xy}^2=(x^0-y^0)^2-|\vecx-\vecy|^2.
\eeq
Examples of such operators have arisen in the causal set theory program
\cite{2007gr.qc.....3099S, Aslanbeigi:2014zva, 2010PhRvL.104r1301B,
2013CQGra..30s5016D, 2014CQGra..31i5007G}.
This operator is clearly linear, real, Poincare-invariant and retarded. 
It is shown in Appendix \ref{ret_Box} that there are choices of 
$a$ and $f$ for which $\widetilde \Box$ is also stable and 
has the desired infrared behaviour \eqref{IRcondition}. 
One such choice is
\beq
f(s)=\frac{4}{\pi}\delta(s-\epsilon)-\frac{e^{-s/2}}{4\pi}(24-12s+s^2), \qquad
a=-2,
\label{mBoxEXfa}
\eeq
where $\epsilon$ is an infinitesimally small positive number.

\begin{figure}
        \centering
        	\includegraphics[width=0.5\textwidth]{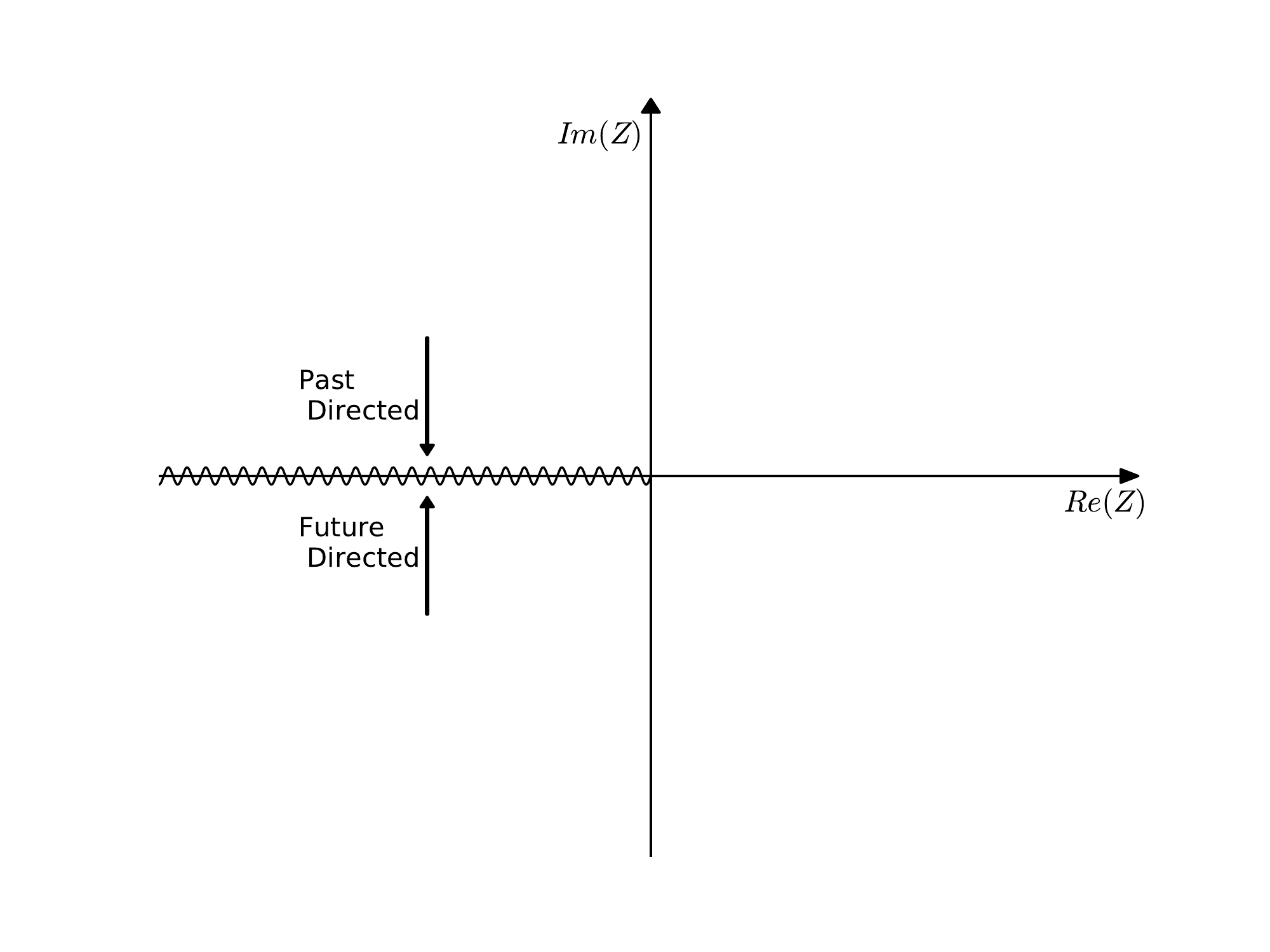}
	 \caption{Analytic structure of $B(p)$ in the complex plane of $Z=p\cdot p/\Lambda^2$}
	\label{analytic}
\end{figure}

\begin{figure}
        \centering
        	\includegraphics[width=0.7\hsize]{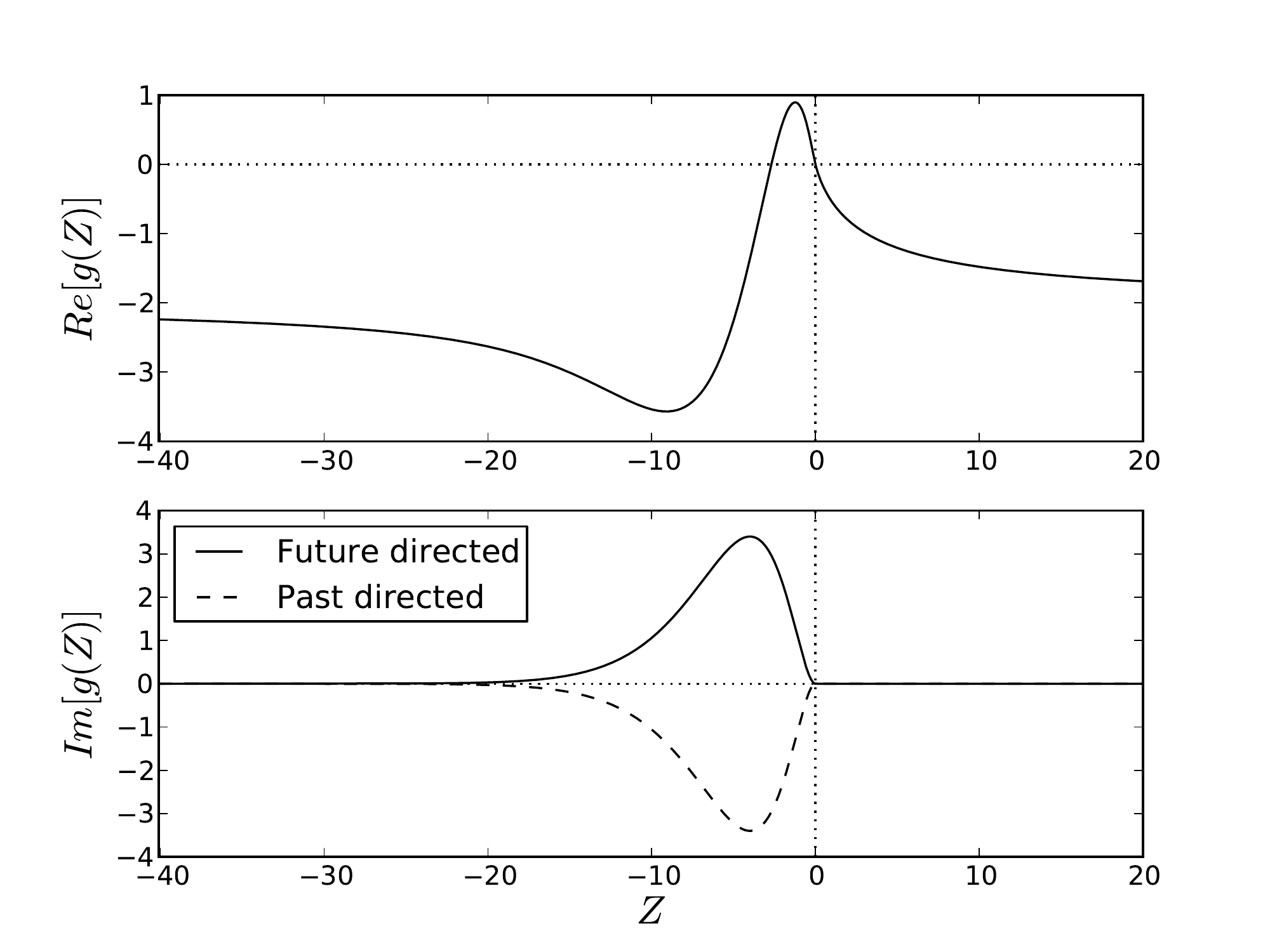}
	 \caption
	 {The Fourier transform $B(p)=g(p\cdot p/\Lambda^2)$ of $\widetilde \Box$ defined in \eqref{mBoxEX},
	  where $a$ and $f$ are given by \eqref{mBoxEXfa}.}
	\label{BPex}
\end{figure}

The eigenvalues $B(p)$ of $\widetilde \Box$ take the form (see \cite{Aslanbeigi:2014zva})
\begin{align}
\Lambda^{-2}B(p)&=\lim_{\epsilon\to0^+}g((p+ip_{\epsilon})\cdot (p+ip_{\epsilon})/\Lambda^2),\label{Bgpp}\\
g(Z)&=a+4\pi Z^{-\frac{1}{2}}\int_{0}^{\infty}f(s^2)s^{2}K_{1}(Z^{1/2}s)ds,
\end{align}
where $p_{\epsilon}$ is an infinitesimally small ($p_{\epsilon}\cdot p_{\epsilon}=-\epsilon^2 $), timelike, 
and future-directed ($p_{\epsilon}^0>0$) wave-vector. The analytic structure of $B(p)$ is shown in 
Figure \ref{analytic}. Figure \ref{BPex} shows the behaviour of $B(p)$ as a function of $p\cdot p$ and
$\text{sgn}(p^0)$ for the choice of $f$ and $a$ given in \eqref{mBoxEXfa}.

%\begin{figure}
%        \centering
%        \begin{subfigure}[t]{0.5\hsize}
%                \includegraphics[trim=50 60 50 40, clip=true, width=\hsize]{branchCutPython}
%                \caption{}
%                \label{analyticpZ}
%        \end{subfigure}%
%        \begin{subfigure}[t]{0.50\hsize}
%                \includegraphics[trim=50 40 50 40, clip=true, width=\hsize]{greenIntg}
%                \caption{}
%                \label{analyticp0}
%        \end{subfigure}
%        \caption{The analytic structure of $B(p)$ in the complex plane of (a) $Z=p\cdot p/\Lambda^2$
%        			and (b) $p^0$. For spacelike momenta ($Z>0$), $B(p)$ is real. For timelike momenta,
%			it is complex with an imaginary part whose sign is opposite for 
%			past-directed and future-directed momenta.}
%        \label{analytic}
%\end{figure}

%==========================================================
\section{Classical Theory}
\label{classicalTheory}
%==========================================================
How would such non-local and retarded evolution manifest itself?
To get a start on answering this question, 
we modify the evolution of a massless scalar field $\phi$
coupled to a source $J(x)$ via
$\Box \rightarrow \widetilde \Box$:
\beq
	\Box\phi(x)=J(x) \rightarrow \widetilde \Box\phi(x)=J(x).
\eeq
It is worth noting that the solutions of
$\widetilde \Box\phi(x)=0$ are identical
to those of $\Box\phi(x)=0$. This follows from requiring a stable 
evolution for $\widetilde \Box$ (see \cite{Aslanbeigi:2014zva}).
As we will see in Section \ref{GFclassical}, however,
the story changes when $J(x)\neq0$.

%-----------------------------------------------------------------------------------------------------
\subsection{Absence of an action principle}
\label{ActionPR}
%-----------------------------------------------------------------------------------------------------
It is natural to ask whether an action principle exists for $\phi$,
whose variation would produce
the non-local equation of motion $\widetilde \Box\phi(x)=J(x)$.
One might propose to substitute $\Box$ with $\widetilde \Box$ 
in the action of a massless scalar field:
\beq
S[\phi]=\int d^4x\left( \frac{1}{2}\phi(x) \widetilde \Box \phi(x) - J(x)\phi(x)\right).
\label{Stilde}
\eeq
Requiring $S[\phi]$ to be stationary with respect to first order variations in
$\phi$ we find
\footnote{
To see this, it is instructive to express the action in Fourier space. Define the Fourier transform
$f(p)$ of $f(x)$ via
\beq
f(x) = \int\frac{d^4p}{(2\pi)^4}f(p)e^{ip\cdot x}.
\eeq
Then, it can be shown that
\beq
S=\int \frac{d^4p}{(2\pi)^4}\left[\phi(p)^* \frac{1}{4}(B(p)+B(p)^*)\phi(p) - \phi(p)^*J(p)\right].
\eeq
Requiring $S$ to be stationary with respect to first order variations $\phi(p)$ we find
\beq
\frac{1}{2}(B(p)+B(p)^*)\phi(p)=J(p).
\eeq
}
\beq
\frac{1}{2}(\widetilde \Box + \widetilde \Box^T)\phi(x)=J(x),
\eeq
where $\widetilde \Box^T$ is defined in Fourier space via
\beq
\widetilde \Box^Te^{ip\cdot x} = B(p)^*e^{ip\cdot x}. 
\eeq
In the case of the retarded operator \eqref{mBoxEX}, for instance, 
$\widetilde \Box^T\phi(x)$ is the right hand side of \eqref{mBoxEX} with the domain
of integration changed to the causal {\it future} of point $x$. 
Therefore, \eqref{Stilde} does not lead to a retarded equation of motion.

Due to the absence of a local Lagrangian description,
quantizing a massless scalar field theory built upon
$\widetilde \Box$ is non-trivial. We shall address this problem
in Section \ref{QuantumFormalism}, where we argue that the 
the Schwinger-Keldysh quantization scheme can still
be used to obtain the desired non-local quantum field theory.

%But this action does not lead to the desired classical limit. One can see this by finding the expectation value of the field
%\beq
%\langle \phi(x)\rangle=\int d^4y G_s(x,y) J(y),
%\eeq
%where $G_s$ is the inverse of $\widetilde \Box_s=\frac{1}{2}(\widetilde \Box+\widetilde \Box^T)$ and $^T$ is transpose operator in time ($\widetilde \Box^{T}$ is an operator with eigenvalues $B^*(p)$). 
%This equation is different from the desired EOM \eqref{EOM2}, as $\widetilde \Box+\widetilde \Box^{T}$ is no longer a retarded operator (in fact, it's time symmetric.) So Feynman path integral does not result in a retarded classical EOM.
%-----------------------------------------------------------------------------------------------------
\subsection{Green's function}
\label{GFclassical}
%-----------------------------------------------------------------------------------------------------
The Green's functions of $\Box$ and $\widetilde \Box$ are quite different,
especially in the ultraviolet (UV) where their spectra differ.
One important difference is that $\widetilde \Box$
, unlike $\Box$, has a unique inverse. Since
$\widetilde \Box$ is a retarded operator by definition, it only has a retarded
Green's function.
Recall that $\Box$ has both a retarded $G^R(x,y)$
and advanced $G^A(x,y)$ Green's function:
\beq
	\Box_x G^{R,A}(x,y)=\delta^{(4)}(x-y),
\eeq
which satisfy the following ``boundary conditions": 
$G^R(x,y)$ vanishes unless $x\succ y$ ($x$ is in the causal future of $y$),
and $G^A(x,y)$ vanishes unless $y\succ x$. 
The two Green's functions are related to one another via $G^A(x,y)=G^R(y,x)$.
In the case of $\widetilde \Box$, Green's function is unique (just the retarded one) and switching the arguments of the retarded Green's function
{\it does not} produce another Green's function. Let us show why this is.

\begin{figure}
        \centering
        	\includegraphics[width=0.7\hsize]{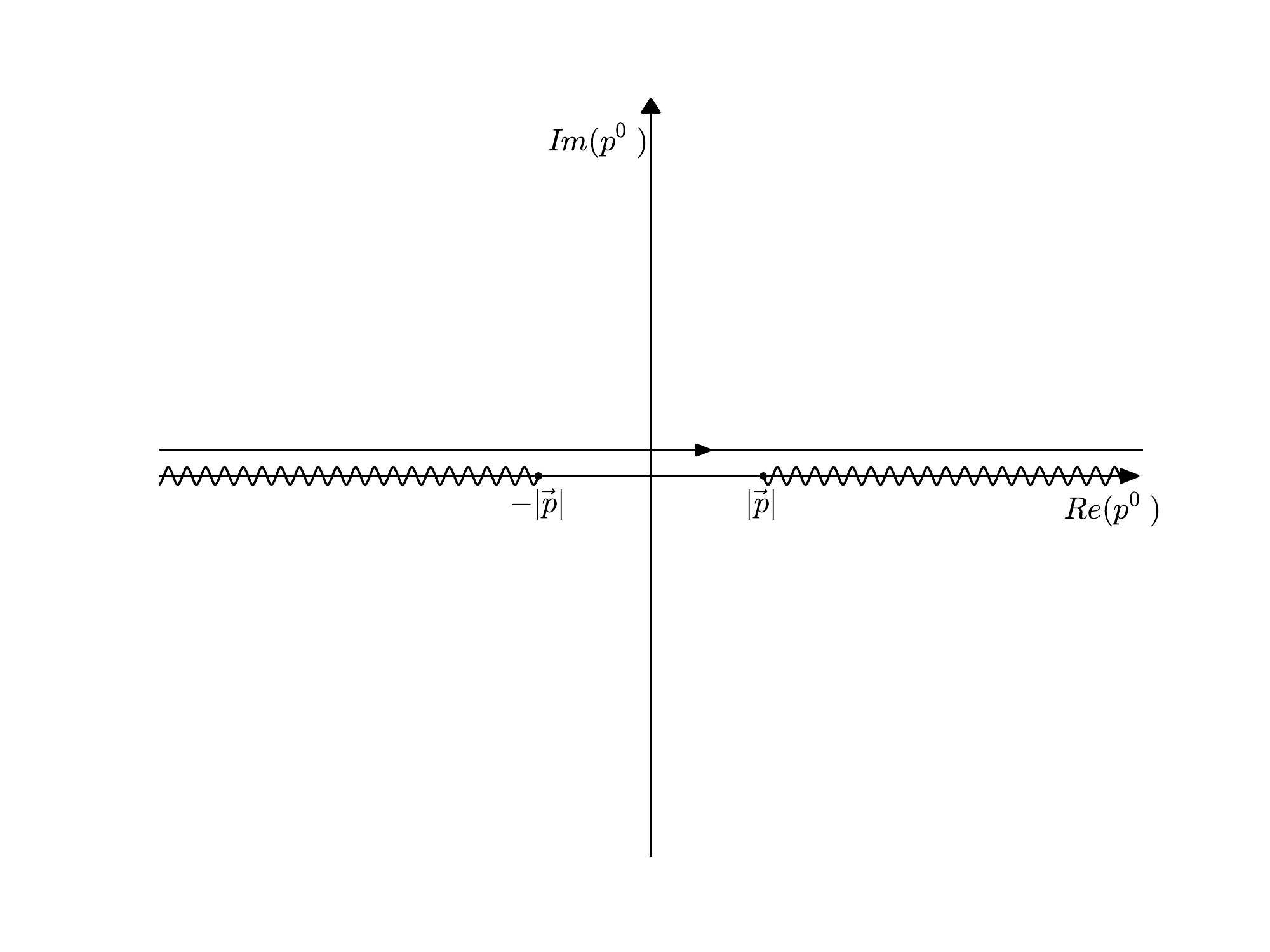}
	 \caption{The integration path in the complex $p^0$ plane which defines the 
	 		retarded Green's function associated with $\widetilde \Box$.}
	\label{analyticGR}
\end{figure}

Let $\widetilde{G}(x,y)$ denote the Green's function 
associated with $\widetilde \Box$:
\beq
	\widetilde \Box_x \widetilde{G}(x,y)=\delta^{(4)}(x-y),
\eeq
Note that $\widetilde{G}(x,y)$ can be expressed as
\beq\label{rgreen}
	\widetilde{G}(x,y)=\int \frac{d^4p}{(2\pi)^4}\frac{1}{B(p)}e^{ip\cdot(x-y)}.
\eeq
The path of integration in the complex $p^0$ plane is shown in
Figure \ref{analyticGR}. % shows the path of integration in the complex $p^0$ plane.
This comes from the fact that $\widetilde \Box$ is a retarded operator, so $B(p)$ analytically continued to the complex $p^0$ plane takes its value above the cut. When $B(p)$ has no zeros in complex plane apart from at $p\cdot p=0$,
which is guaranteed by the stability requirement, 
this choice of contour ensures that $\widetilde{G}(x,y)\equiv \widetilde{G}^R(x,y)$ is indeed retarded.
Switching the arguments of $\widetilde{G}^R(x,y)$, we find
\begin{align}
	\widetilde{G}^R(y,x)&=\int \frac{d^4p}{(2\pi)^4}\frac{1}{B(p)}e^{ip\cdot(y-x)}\\
		     		&=\int \frac{d^4p}{(2\pi)^4}\frac{1}{B(-p)}e^{ip\cdot(x-y)}\\
		     		&=\int \frac{d^4p}{(2\pi)^4}\frac{1}{B(p)^*}e^{ip\cdot(x-y)},\label{rgreenConj}
\end{align}
where in the second line we have changed integration variables from $p$ to $-p$.
Then
\beq
	\widetilde \Box_x \widetilde{G}^R(y,x) = \int \frac{d^4p}{(2\pi)^4}\frac{B(p)}{B(p)^*}e^{ip\cdot(x-y)} \neq \delta^{(4)}(x-y),\label{GAmBox}
\eeq
since $B(p)$ is generically complex. As we will see in the sections to come, 
the fact that $\widetilde \Box$ has a unique inverse plays a crucial role in the quantum 
theory of $\widetilde \Box$.

%==========================================================
\section{Quantum Theory}
\label{QuantumFormalism}
%==========================================================
We wish to construct a quantum theory 
of a massless scalar field $\phi$ whose classical limit
reproduces the retarded evolution induced by $\widetilde \Box$.
The quantization scheme which we believe is most suited
in this case is the Schwinger-Keldysh (or double path integral) formalism.
In what follows, we will first review the usual paths to quantization 
(i.e. canonical quantization and the Feynman path integral)
and show why they fail in the case of a non-local and retarded 
operator like $\widetilde \Box$. The goal of these discussions is to make
clear why we choose the Schwinger-Keldysh formalism to construct a
quantum field theory based on $\widetilde \Box$.

%-----------------------------------------------------------------------------------------------------
\subsection{Canonical quantization}
%-----------------------------------------------------------------------------------------------------
Let us consider the canonical quantization of a free massless scalar field $\phi$.
The typical route to quantization is as follows: start from an action principal for $\phi$, 
derive the Hamiltonian in terms of $\phi$ and its conjugate momentum, 
impose equal-time commutation relations, and finally
specify the dynamics via the Heisenberg equation.
There is an equivalent approach, however, which defines 
the theory with no reference to an action principle,
using the Klein-Gordon equation supplemented by 
the so-called Peierls form of the commutation relations:
\begin{align}
	\Box\widehat{\phi}(x)&=0\label{KGeom}\\
	[\widehat{\phi}(x), \widehat{\phi}(y)]&=i\Delta(x,y),\label{Peierls}
\end{align}
where $\Delta(x,y)$ is the Pauli-Jordan function:
\begin{align}
	\Delta(x,y) &= G^R(x,y) - G^A(x,y)\notag\\
			&= G^R(x,y) - G^R(y,x).\label{Pauli-Jordan}
\end{align}
It is well known that \eqref{Peierls} is entirely equivalent to, but more explicitly covariant than,
the more commonly seen equal-time commutation relations 
(see e.g. Section C.2 of \cite{2014PhDT........44A}).
Since $\Delta(x,y)$ is the difference of two Green's functions, 
it satisfies the equation of motion:
\beq
	\Box_x\Delta(x,y) = 0.
\eeq
This is why \eqref{KGeom} and \eqref{Peierls} are consistent with one another:
both the left and right hand side of \eqref{Peierls} vanish when $\Box_x$ is
applied.

It is tempting to build the quantum theory of $\widetilde \Box$ in a similar fashion:
\begin{align}
	\widetilde \Box\widehat{\phi}(x)&=0\label{KGeomNL}\\
	[\widehat{\phi}(x), \widehat{\phi}(y)]&=i\widetilde{\Delta}(x,y)\equiv i(\widetilde{G}^R(x,y) - \widetilde{G}^R(y,x)).\label{PeierlsNL}
\end{align}
In this case, however, $\widetilde{\Delta}(x,y)$ does not satisfy the equation of
motion ($\widetilde \Box_x\widetilde{\Delta}(x,y)\neq0$)
because $\widetilde{G}^R(y,x)$ is not a Green's function of $\widetilde \Box$
(see Section \ref{classicalTheory} and \eqref{GAmBox}).
Therefore, the equation of motion \eqref{KGeomNL} is not consistent with the
commutation relations \eqref{PeierlsNL}.

It is worth noting that the root of this inconsistency is that the Fourier transform $B(p)$
of $\widetilde \Box$ is complex, which in turn follows from the fact that $\widetilde \Box$
is retarded by definition. 
In Section \ref{SchKel} we will arrive at a consistent quantum theory via the
Schwinger-Keldysh formalism, using which we also build a Hilbert space
representation of the theory. There we will see that the equation of motion
\eqref{KGeomNL} is given up in favour of the commutation relations 
\eqref{PeierlsNL}. As it turns out, the degree to which \eqref{KGeomNL} is violated
depends on the imaginary part of $B(p)$.

%-----------------------------------------------------------------------------------------------------
\subsection{Feynman path integral}
%-----------------------------------------------------------------------------------------------------
The Feynman path integral formalism requires a local Lagrangian description
for the scalar field $\phi$. As was argued in Section \ref{ActionPR}, however,
this is not viable if one requires a retarded equation of 
motion. Therefore, the Feynman path integral formalism is also not suitable for
quantizing this theory.
%-----------------------------------------------------------------------------------------------------
\subsection{Schwinger-Keldysh formalism}
\label{SchKel}
%-----------------------------------------------------------------------------------------------------
The Schwinger-Keldysh formalism 
has a natural way of incorporating a retarded operator. 
In this approach an amplitude (called the decoherence functional $\mathcal{D}(\phi^+,\phi^-)$) 
is assigned to a pair of paths ($\phi^+,\phi^-$), 
which are constrained to meet at the final time ($\phi^+(t_f,\bold x)=\phi^-(t_f,\bold x)$).
The decoherence functional for a free massless scalar field takes the form
\beq\label{decoher}
\mathcal{D}(\phi^+,\phi^-)={\rm Exp}\left[ i \int d^4x \frac{1}{2}\phi^q \Box^R \phi^{cl}+\frac{1}{2}\phi^{cl} \Box^A \phi^q+\frac{1}{2}\phi^q \Box^K \phi^q\right],
\eeq
where 
\bea
\phi^{cl} \equiv \frac{1}{\sqrt{2}}\left(\phi^++\phi^-\right),\\
\phi^{q} \equiv \frac{1}{\sqrt{2}}\left(\phi^+-\phi^-\right).
\eea
In \eqref{decoher}, $\Box^R$ is the retarded d'Alembertian,
$\Box^A=(\Box^R)^{\dag}$ is the advanced d'Alembertian, and $\Box^K$ is an anti-Hermitian operator 
which contains information about the initial wave function \cite{kamenev2005course}.%
\footnote{
The retarded and advanced d'Alembertians are defined via $G^{R,A}(\Box^{R,A}f)=f$ for all suitable test functions $f$, where
$G^{R,A}$ are the integral operators associated with the retarded and advanced Green's functions $G^{R,A}(x,y)$.
}
Any source term $J(x)$ 
can be included by adding $-J\phi^++J\phi^-=-\sqrt{2}J\phi^q$ to the integrand.

Any $n$-point function in this theory is given by
\bea
\langle&& \phi^{(\alpha_1)}(x_1)\cdots \phi^{(\alpha_n)}(x_n)\rangle\notag\\
&&=\int D\phi^+D\phi^{-}\phi^{(\alpha_1)}(x_1)\cdots\phi^{(\alpha_n)}(x_n)\mathcal{D}(\phi^+,\phi^-),
\eea
where $\alpha_i \in \{+,-,q,cl\}$. These correlation functions are related to the correlation functions in Hilbert space representation by the following rule:
\bea\label{SchKel_to_Hilbert}
\langle&& \phi^+(x_1)\cdots \phi^+(x_n)\phi^-(y_1)\cdots \phi^-(y_m)\rangle\notag\\
&&=\langle0|\widetilde T\left[\widehat \phi(y_1)\cdots \widehat \phi(y_m)\right]T\left[\widehat \phi(x_1)\cdots \widehat \phi(x_n)\right]|0\rangle
\eea
where $T(\widetilde T)$ is the (anti) time-ordered operator, and $|0\rangle$ is the vacuum state of the free theory.

In order to come up with a quantum theory for a non-local retarded operator, we replace $\Box^R$ with $\widetilde \Box$ in \eqref{decoher} (and $\Box^K$ with $\widetilde \Box^K$\footnote{We still need to determine $\widetilde \Box^K$. This has been done in \ref{GreenK}.}).
%.........................................................................................................................
\subsubsection{Classical limit}
%.........................................................................................................................
%Here, we show that double path integral has the right classical limit, i.e. it describes the evolution \eqref{EOM2}. Variation of the action with respect to $\phi^{cl}$ results in
%\beq
%B^A\phi^q=0.
%\eeq
%Since $\phi^q$ must vanish at final time, solution to this equation is $\phi^q=0$. Variation with respect to $\phi^q$ (at $\phi^q=0$) yields
%\beq
%2B^R \phi^{cl}=\sqrt{2}J(x).
%\eeq
%Note that for $\phi^q=0$, $\phi\equiv \phi^+=\phi^-=\sqrt{2}\phi^{cl}$. As a result, we get
%\beq
%B^R \phi=J(x).
%\eeq

Before going any further, let us take a look at the classical limit of this theory. Performing Gaussian integrals (in the presence of a source term), we get
\bea
\langle \phi^{cl}(x)\rangle&=& \frac{1}{\sqrt{2}}\int d^4y~\widetilde G^R(x,y)J(y),\\
\langle \phi^{q}(x)\rangle&=& 0,
\eea
resulting in
\beq
\langle \phi^{+}(x)\rangle=\langle \phi^{-}(x)\rangle= \int d^4y~\widetilde G^R(x,y)J(y).
\eeq
It shows that in the classical limit where the field is represented by its expectation value, there is no difference between $\phi^+$ and $\phi^-$ and both satisfy the retarded equation of motion $\widetilde \Box \phi=J$.

%.........................................................................................................................
\subsubsection{Green's functions}
%.........................................................................................................................
Let us consider the two point correlation functions of this theory in the absence of any source

\bea
-i\langle&& \phi^{cl}(x)\phi^q(y)\rangle = \widetilde G^R(x,y)\\
-i\langle&& \phi^{q}(x)\phi^{cl}(y)\rangle\equiv \widetilde G^A(x,y)=\widetilde G^R(y,x)\\
-i\langle&& \phi^{cl}(x)\phi^{cl}(y)\rangle\equiv\widetilde G^K(x,y)\notag\\
&&=-\int d^4zd^4w~ \widetilde G^R(x,z)\widetilde B^K(z,w)\widetilde G^A(w,y)\\
-i\langle&& \phi^{q}(x)\phi^{q}(y)\rangle=0
\eea
where $\widetilde B^K(x,y)$ is the kernel of $\widetilde \Box^K$\footnote{If $\delta_y(x)\equiv \delta^{(4)}(x-y)$, then $\widetilde B^K(x,y)\equiv (\widetilde \Box^K\delta_y)(x)$. With this definition, $(\widetilde \Box^K\phi)(x)=\int d^4y \widetilde B^K(x,y)\phi(y)$.}. Using the definition of $\phi^q$ and $\phi^{cl}$, we get
\beq
-i\langle \phi^+(x)\phi^+(y)\rangle=\frac{1}{2}\left[\widetilde G^K(x,y)+\widetilde G^R(x,y)+\widetilde G^A(x,y)\right],\label{Tfunction}
\eeq
\beq
-i\langle \phi^-(x)\phi^-(y)\rangle=\frac{1}{2}\left[\widetilde G^K(x,y)-\widetilde G^R(x,y)-\widetilde G^A(x,y)\right],\label{ATfunction}
\eeq
\beq
-i\langle \phi^-(x)\phi^+(y)\rangle=\frac{1}{2}\left[\widetilde G^K(x,y)+\widetilde G^R(x,y)-\widetilde G^A(x,y)\right].\label{eq9}
\eeq
Note that if this theory has an equivalent representation in terms of field operator in a Hilbert space, then the above mentioned terms correspond to time-ordered two point function, anti time-ordered two point function and two point function respectively (see \eqref{SchKel_to_Hilbert}).

We require that the theory describes a free scalar field in flat space-time at its ground state. As a result, all $n$-point correlation functions of this theory must be translation invariant,
\beq
\langle \phi^{(\alpha_1)}(x_1)\cdots \phi^{(\alpha_n)}(x_n)\rangle=\langle \phi^{(\alpha_1)}(x_1+y)\cdots \phi^{(\alpha_n)}(x_n+y)\rangle.
\eeq
This condition requires that all operators $\widetilde \Box$, $\widetilde \Box^{\dagger}$ and $\widetilde \Box^K$ must be translation invariant. Consequently, we get
\bea
\widetilde \Box^K e^{ip\cdot x}&=&\widetilde B^K(p)e^{ip\cdot x},\\
\widetilde G^K(x,y)&=&-\int \frac{d^4p}{(2\pi)^4}\widetilde G^R(p)\widetilde B^K(p)\widetilde G^A(p)e^{ip\cdot (x-y)}~~
\eea
Note that $\widetilde \Box^K$ is an anti-Hermitian operator. It means $\widetilde B^K(p)$ is a total imaginary number (and $\widetilde G^K(p)\equiv -\widetilde G^R(p)\widetilde B^K(p)\widetilde G^A(p)$ is also total imaginary since $\widetilde G^R(p) \widetilde G^A(p)$ is real.) 

%.........................................................................................................................
\subsubsection{Fixing $\widetilde G^K$}\label{GreenK}
%.........................................................................................................................

From here on, we assume that there is a Hilbert space representation of this theory with a Hamiltonian evolution. We will justify this assumption later by finding the representation itself. In Appendix \ref{AppFDT} we show that this assumption leads to the following relation, when the quantum system is in its ground state
\beq
\widetilde G^K(p)={\rm sgn}(p^0)\left[\widetilde G^R(p)-\widetilde G^A(p)\right].\label{FDT}
\eeq
Note that \eqref{FDT} is nothing but the fluctuation dissipation theorem (FDT) at zero temperature. This fixes the eigenvalues of $\widetilde \Box^K$ as follows:
\beq
\widetilde B^K(p)=2i{\rm Im}B(p){\rm sgn(p^0)}
\eeq

%.........................................................................................................................
\subsubsection{Hilbert space representation}
\label{Hilbert_representation}
%.........................................................................................................................

We wish to find an equivalent Hilbert space representation in terms of a field operator $\widehat \phi(x)$ for this theory. 
As we mentioned earlier, \eqref{eq9} is the two point function of such a representation,
\beq
W(x,y)\equiv \langle 0|\widehat \phi(x)\widehat \phi(y)|0\rangle=\langle \phi^-(x)\phi^+(y)\rangle,
\eeq
where $|0\rangle$ is the ground state. If we use \eqref{eq9} and \eqref{FDT}, we arrive at  
\beq
W(x,y)=\int \frac{d^4p}{(2\pi)^4}\frac{2{\rm Im}[B(p)]\theta(p^0)}{|B(p)|^2}e^{ip\cdot(x-y)},
\label{TPF}
\eeq
where we call $\widetilde W(p)\equiv \frac{2{\rm Im}[B(p)]\theta(p^0)}{|B(p)|^2}$. Since $W(x,y)$ is a positive operator,  
$\text{Im}[B(p)]\theta(p^0)$ must be a non-negative number. So, we further {\it assume}
\beq\label{sgn}
{\rm sgn}\left({\rm Im}[B(p)]\right)={\rm sgn}(p^0).
\eeq

Once this condition is satisfied, the field operator $\widehat \phi(x)$ and ground state $|0\rangle$, defined to be 
\bea
\widehat \phi(x)&=&\int \frac{d^4p}{(2\pi)^2}\sqrt{\widetilde W(p)}\left(\widehat a_pe^{ip\cdot x}+\widehat a_p^{\dagger} e^{-ip\cdot x}\right),\label{eq10}\\
\left[\widehat a_p, \widehat a_q\right] &=& \delta^{(4)}(p-q),\\
\widehat a_p|0\rangle&=&0~~ \forall p,\label{groundstate}
\eea
yield the desired correlation functions.

Note that $a_p$ is only defined for time-like future-directed $p$, because otherwise $\widetilde W(p)$ is zero in the field expansion. It means that all time-like future-directed (positive energy) momenta contribute to the field expansion \eqref{eq10}. 

%.........................................................................................................................
\subsubsection{Hamiltonian}
%.........................................................................................................................

By definition, time evolution operator is the operator that evolves $\widehat \phi(x)$ in time,
\beq
\widehat \phi(t,\bold x)=\widehat U(t,t_0)\widehat \phi(t_0,\bold x)\widehat U^{\dagger}(t,t_0).
\eeq
It can be directly checked that 
\bea
\widehat U(t,t_0)&=&e^{-i\widehat H_0(t-t_0)},\\
\widehat H_0&=&\int d^4p~p^0\widehat a_p^{\dagger}\widehat a_p,
\eea
gives the right time evolution. 

State $|0\rangle$ defined in \eqref{groundstate} is the {\it ground state} of this Hamiltonian. Excited states ($n$-particle states) can be built by acting $a^{\dagger}$'s on $|0\rangle$,
\beq
|p_1\cdots p_n\rangle=\widehat a^{\dagger}_{p_1}\cdots \widehat a^{\dagger}_{p_n}|0\rangle.
\eeq
The excited state $|p\rangle$ represents a particle with energy $p^0$ and momentum $\bold p$\footnote{Momentum operator $\widehat {\bold P}\equiv \int d^4p~\bold p~\widehat a_p^{\dagger}\widehat a_p$ is the generator of spacial translation.} where $p^0$ is independent of $\bold p$\footnote{Note that these states are different from the usual states $|\bold p\rangle$ used in LQFT which describe a particle with momentum $\bold p$ and energy $|\bold p|$.}. This shows that the theory contains a continuum of massive particles with positive energy. The existence of a continuum of massive particles in the context of Causal Set theory also has been pointed out in \cite{Belenchia:2014fda}, although their result is rather different in some other aspects.

%.........................................................................................................................
\subsubsection{Comparison to local evolution}
%.........................................................................................................................

At this point, it would be illustrative to consider the result of this formalism for LQFT. In this case
\beq
B(p)=B_{local}(p)=(p^0+i\epsilon)^2-|\bold p|^2,
\eeq
where $\epsilon$ is a small positive number taken to zero at the end of calculation.
The two point function is given by 
\beq
\widetilde W(p)=2\frac{\epsilon p^0}{(p^2)^2+(\epsilon p^0)^2}\theta(p^0)=2\pi\delta (p^2)\theta(p^0).
\eeq
As a result, 
\bea
&W(x,y)=\int \frac{d^4p}{(2\pi)^4}2\pi \delta(p^2)\theta(p^0)e^{ip\cdot(x-y)},\\
&\widehat \phi(x)=\int\frac{d^4p}{(2\pi)^2}\sqrt{2\pi\delta(p^2)\theta(p^0)}\left(\widehat a_p e^{ip\cdot x}+\widehat a^{\dagger}_p e^{-ip\cdot x}\right).\label{eq12}
\eea
Two point function and field expansion are exactly the ones we expected. 
Only on-shell particles ($p\cdot p=0$) contribute to the field expansion. 

Here, we see one important difference between local and retarded non-local evolution. 
In the local case, only on-shell modes ($p\cdot p=0$) contribute to the field expansion. 
As a result, excited states of the theory consist of all {\it on-shell} particles. In non-local 
retarded case (where generically ${\rm Im}[B(p)] \neq 0$), off-shell modes ($p\cdot p\neq 0$) also 
contribute to the field expansion. 
Consequently, one expect the existence of off-shell modes in "in" and "out" state of scatterings in the interacting theory.

Let us investigate properties of $\widetilde W(p)$ for a generic non-local retarded operator. 
First of all, it is only non-zero for time-like future-directed momenta. 
This means that only time-like future-directed momenta contribute to the field expansion 
and can exist in "in" and "out" state (particles with time-like momentum and positive energy).

Considering that $B(p)$ is only zero at $p\cdot p=0$, $\widetilde W(p)$ is a finite number for 
all $p\cdot p \neq 0$ (we will see the significance of this result in \ref{cross section}). 
On the other hand, since in the subspace of on-shell modes $\widetilde \Box$ operator is exactly the same as $\Box$, 
we conclude that  $\widetilde W(p)=2\pi\delta(p^2)\theta(p^0)$ for $p\cdot p=0$. 
Therefore, $\widetilde W(p)$ consists of a divergent part at $p\cdot p=0$ and a finite part for $p\cdot p \neq0$. 
This means that there are two different contributions to the field expansion \eqref{eq10}, one from on-shell 
modes that is the same as \eqref{eq12} and one from off-shell modes which only exists in 
the case of non-local retarded evolution
\bea
\widehat \phi(x)&&=\int\frac{d^4p}{(2\pi)^2}\sqrt{2\pi\delta(p^2)\theta(p^0)}\left(\widehat a_p e^{ip\cdot x}+\widehat a^{\dagger}_p e^{-ip\cdot x}\right)\notag\\
&&+\int_{p^2\neq0}\frac{d^4p}{(2\pi)^2}\sqrt{\widetilde W(p)}\left(\widehat a_p e^{ip\cdot x}+\widehat a^{\dagger}_p e^{-ip\cdot x}\right).\label{field_operator}
\eea

%-----------------------------------------------------------------------------------------------------
\subsection{Sorkin--Johnston quantization} 
%-----------------------------------------------------------------------------------------------------
The Sorkin-Johnston (SJ) proposal defines a unique vacuum state for a free massive 
scalar field
in an arbitrarily curved spacetime \cite{2012JHEP...08..137A}. 
This proposal is a continuum generalization of Johnston's formulation of
a free quantum scalar field theory on a background causal set \cite{Johnston}.
As is the case for $\widetilde \Box$, canonical quantization does not admit an obvious
generalization for a causal set. The SJ quantization scheme
uses only the retarded Green's function $G_R(x,y)$ to arrive at the quantum theory.
Since $\widetilde \Box$ also admits a retarded Green's function, 
one can apply the SJ prescription to arrive at a free quantum field theory
of the massless scalar field we have been considering. 
In what follows, we will show that the SJ proposal applied to $\widetilde \Box$
produces the same free quantum theory as the Schwinger Keldysh formalism,
provided condition \eqref{sgn} is met.

Consider the corresponding
integral operator of the kernel $i\Delta(x,y)=G_R(x,y)-G_R(y,x)$:
\beq
(i\Delta f)(x)=\int i\Delta(x,y)f(y)d^4y.
\eeq
It can be shown that $i\Delta$ is Hermitian, which implies it has real eigenvalues, 
and that its non-zero eigenvalues come in positive and negative pairs:
\beq
(i\Delta T_{\vecp})(x)=\lambda_{\vecp}^2T_{\vecp}(x)
\qquad\rightarrow\qquad
(i\Delta T_{\vecp}^*)(x)=-\lambda_{\vecp}^2T^*_{\vecp}(x).
\eeq
We have assumed here that the eigenfunctions $T_{\vecp}$ form an orthonormal basis of $L^2$,
which can always be achieved since $i\Delta$ is Hermitian.
The Sorkin-Johnston proposal is then to define the two-point function to be the
positive part of $i\Delta(x,y)$ in the following sense:
\beq
\langle0|\widehat{\phi}(x)\widehat{\phi}(y)|0\rangle = \sum_{\vecp}\lambda_{\vecp}^2T_{\vecp}(x)T^*_{\vecp}(y).
\eeq

Taking $G_R(x,y)$ to be the retarded Green's function of $\widetilde \Box$ (see \eqref{rgreen} and \eqref{rgreenConj}), we find
\beq
i\Delta e^{ip\cdot x} = \frac{2\text{Im}(B(p))}{|B(p)|^2}e^{ip\cdot x},
\eeq
which using the SJ formalism then leads to the two-point function
\beq
\langle0|\widehat{\phi}(x)\widehat{\phi}(y)|0\rangle = \int \frac{d^4p}{(2\pi)^4}\frac{2\text{Im}(B(p))}{|B(p)|^2}\theta(\text{Im}(B(p)))e^{ip\cdot x}.
\eeq
If condition \eqref{sgn} is satisfied, this two-point function is at that derived from the
Schwinger-Keldysh formalism (see \eqref{TPF} and \eqref{sgn}). It is reassuring that two
different paths to quantization, at least at the free level, lead to the same theory.

%%%%%%%%%%%%%%%%%%%%%%%%%%%%%%%%%%%%%%%
%%%%%%%%%%%%%%%%%%%%%%%%%%%%%%%%%%%%%%%
%%%%%%%%%%%%%%Interacting Theory%%%%%%%%%%%%%%%%
%%%%%%%%%%%%%%%%%%%%%%%%%%%%%%%%%%%%%%%
%%%%%%%%%%%%%%%%%%%%%%%%%%%%%%%%%%%%%%%

%%%%%%%%%%%%%%%%%%%%%%%%%%%%%%%%
\section{Interacting Field Theory}
\label{interaction}
%%%%%%%%%%%%%%%%%%%%%%%%%%%%%%%%
Let us now consider the interacting theory. We introduce the interaction in the Hilbert space representation by adding a potential term to the free Hamiltonian as follows:
\beq
\widehat H(t)=\widehat H_0+\int d^3\bold x V(\widehat \phi(t,\bold x)).
\eeq
Starting with a general initial wave function, one is able to find the final state of the system by solving Heisenberg equation of motion in principle. However, in practice this is a very hard task to do. So, we try to find the S-matrix amplitudes perturbatively. 

In order to do so, we can use the available machinery of LQFT, and move to the interaction picture. Time evolution in the interaction picture is given by
\beq
\widehat U_I=T e^{-i\int d^4 x V(\widehat \phi_I)}
\eeq
where $\widehat \phi_I$ is the field in the interaction picture given by \eqref{eq10}. Perturbative expansion of $\widehat U_I$ yields S-matrix amplitudes. Performing the calculations to find the S-matrix, we come up with modified Feynman rules for this theory. We explain these modifications in the following two examples.  

%%%%%%%%%%%%%%%%%%%%%%%%%%%%%%%%
\subsection{Example 1: 2-2 Scattering $p_1p_2 \to q_1q_2$ in $\frac{\lambda}{4!}\phi^4$ theory}
%%%%%%%%%%%%%%%%%%%%%%%%%%%%%%%%

Scattering amplitude $S_{q_1q_2,p_1p_2}$ is given by
\beq
S_{q_1q_2,p_1p_2}=\langle q_1q_2|T e^{-i\int d^4 x \frac{\lambda}{4!}\widehat \phi^4_I}|p_1p_2\rangle.
\eeq
To first order in $\lambda$, it yields
\bea
&S_{q_1q_2,p_1p_2}=-i\frac{\lambda}{4!} \int d^4x~\langle q_1q_2| \widehat \phi^4_I(x) |p_1p_2\rangle\notag\\
&=\frac{-i\lambda}{(2\pi)^4}\sqrt{\widetilde W(p_1)\widetilde W(p_2)\widetilde W(q_1)\widetilde W(q_2)}\delta^{(4)}(\sum p -\sum q),~~~~~\label{Sphi4}
\eea
where we have substituted for $\widehat \phi_I$ from \eqref{eq10}. It is interesting to note that \eqref{Sphi4} is time reversal invariant. 

In the transition from local to retarded non-local propagation, here we see the first change in the scattering amplitudes. The values assigned to each external line have changed from $\sqrt{2\pi \delta(p^2)\theta(p^0)}$ to $\sqrt{\widetilde W(p)}$. Note that here the scattering amplitude is computed in the basis of 4-momentum $|p\rangle$ which is different from 3-momentum basis $|\bold p\rangle$ of LQFT.

%%%%%%%%%%%%%%%%%%%%%%%%%%%%%%%%
\subsection{Example 2: 2-2 Scattering $p_1p_2\rightarrow q_1q_2$ in $\frac{\lambda}{3!}\phi^3$ theory}
%%%%%%%%%%%%%%%%%%%%%%%%%%%%%%%%

In this case, $S_{q_1q_2,p_1p_2}$ is given by
\beq
S_{q_1q_2,p_1p_2}=\langle q_1q_2|T e^{-i\int d^4 x \frac{\lambda}{3!}\widehat \phi^3_I}|p_1p_2\rangle.
\eeq
To second order in $\lambda$, it yields
\bea
&S_{q_1q_2,p_1p_2}=\frac{1}{2}(\frac{-i\lambda}{3!})^2 \int d^4x~d^4y~\langle q_1q_2| T\widehat \phi^3_I(x)\widehat \phi^3_I(y) |p_1p_2\rangle\notag\\
&=\frac{-i\lambda^2}{(2\pi)^8}\sqrt{\widetilde W(p_1)\widetilde W(p_2)\widetilde W(q_1)\widetilde W(q_2)}\delta^{(4)}(\sum p-\sum q)\notag\\
&\times \left[\widetilde G^F(p_1+p_2)+\widetilde G^F(p_1-q_1)+\widetilde G^F(p_1-q_2)\right].\notag
\eea
$\widetilde G^F(p)=\frac{\theta(p^0)}{B(p)}+\frac{\theta(-p^0)}{B^*(p)}$ is the time-ordered two point function \eqref{Tfunction} in Fourier space. In the transition from local to non-local operator, here we see another change in the scattering amplitude. The values assigned to each internal line have changed to the new value for the Feynman propagator $\widetilde G_F(p)$.

\begin{figure}
\centering
\includegraphics[width=0.5\textwidth]{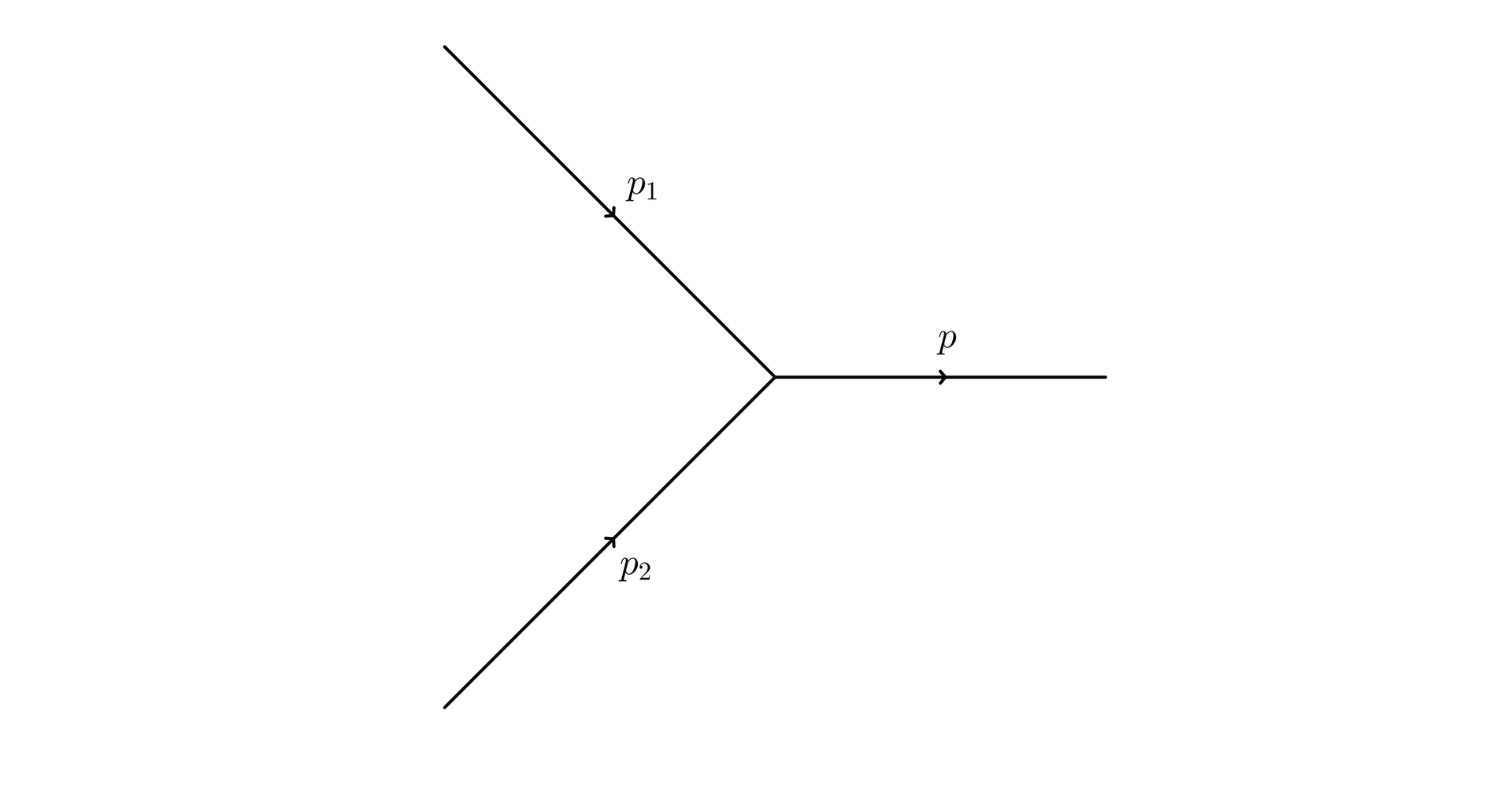}
\caption{The amplitude of this diagram in LQFT is zero, because of the energy momentum conservation; two massless particles cannot produce a massless particle. However, in our theory there is a continuum of massive particles and the amplitude of this scattering is generically non-zero.}
\label{2to1}
\end{figure}

From these examples, it is obvious how scattering amplitudes can be computed in this theory. For any Feynman diagram only the values assigned to external lines and internal lines have changed.
Note that the amplitude of some diagrams in LQFT is zero, as a result of energy-momentum conservation, while in this theory they are not. For example in LQFT $\lambda \phi^3$ theory, the amplitude assigned to diagram \ref{2to1} is zero, because the sum of two (non-parallel) null vectors cannot be a null vector. However, in this theory there is a continuum of massive particles, and for example two on-shell particles can interact and produce one off-shell particle.

%%%%%%%%%%%%%%%%%%%%%%%%%%%%%%%%%%%%%%%
%%%%%%%%%%%%%%%%%%%%%%%%%%%%%%%%%%%%%%%
%%%%%%%%%%%%%%Cross Section%%%%%%%%%%%%%%%%%%
%%%%%%%%%%%%%%%%%%%%%%%%%%%%%%%%%%%%%%%
%%%%%%%%%%%%%%%%%%%%%%%%%%%%%%%%%%%%%%%

%%%%%%%%%%%%%%%%%%%%%%%%%%%%%%%%
\section{From Scattering Amplitude to Transition Rate}
\label{cross-section}
%%%%%%%%%%%%%%%%%%%%%%%%%%%%%%%%
At this point, we want to find the rate of a process using the S-matrix amplitudes. 
In \ref{cross section} we have shown that if one (or more) of the incoming particles is off-shell, then the differential transition rate of such scattering is zero. It means that in order to have a non-zero transition rate (and cross-section), all of the incoming particles must be on-shell. This is the most distinctive property of off-shell particles: cross-section of any scattering with off-shell particles is zero. 

For now consider the scattering from state $|\alpha\rangle=|p_1\cdots p_{N_{\alpha}}\rangle$ to $|\beta\rangle=|q_1\cdots q_{N_{\beta}}\rangle$ where all the incoming particles are on-shell, $p_i^2=0$. Assuming that the interactions happen inside a box with volume $V$ (see \cite{weinberg1995quantum}), differential transition rate is given by
\bea
d\Gamma =&&2\pi^{N_{\alpha}+1} \left[\frac{(2\pi)^3}{V}\right]^{N_{\alpha}-1}\frac{1}{E_{\mathbf{ p_1}}\cdots E_{\mathbf{ p_{N_{\alpha}}}}}\delta^{(4)}(\sum p_i-\sum q_i)\notag\\
&&\times |\widetilde M_{\beta \alpha}|^2 d^4q_1\cdots d^4q_{N_{\beta}},
\eea
where $E_{\mathbf{ p_i}}=|\mathbf{ p_i}|$ and
\beq\label{def_tildeM}
S_{\beta \alpha}=-2\pi i \delta^{(4)}(\sum p_i-\sum q_i)\sqrt{\widetilde W(p_1)\cdots \widetilde W(p_{N_{\alpha}})}\widetilde M_{\beta \alpha}.
\eeq
In the case of 2-2 scattering, the differential cross section is given by
\beq
d\sigma=\frac{d\Gamma}{\frac{u}{V}}=\frac{\pi^2(2\pi)^4}{E_{\mathbf{ p_1}}E_{\mathbf{ p_2}}u}\delta^{(4)}(\sum p_i-\sum q_i)|\widetilde M_{\beta \alpha}|^2d^4q_1d^4q_2,
\eeq
where 
\beq\label{speed}
u=\frac{\sqrt{(p_1.p_2)^2-p_1^2p_2^2}}{p_1^0p_2^0}
\eeq
is the speed of particle 1 in the frame of reference of particle 2 (and vice versa) and $\frac{u}{V}$ is the flux of incoming particles.

%Here, we use the same technique that has been used in Weinberg. Let's assume that all interactions are happening inside a space-time box $T\times L\times L\times L~ (V\equiv L^3)$. This allows for the normalization of states. Following the steps of Weinberg, we get the following differential transition rate from state $|\alpha\rangle=|p_1\cdots p_{N_{\alpha}}\rangle$ to $|\beta\rangle=|q_1\cdots q_{N_{\beta}}\rangle$, 
%\beq
%d\Gamma =2\pi \left[\frac{(2\pi)^3}{V}\right]^{N_{\alpha}-1}\left[\frac{2\pi}{T}\right]^{N_{\alpha}}\delta^{(4)}(\sum p_i-\sum q_i)|M_{\beta \alpha}|^2 d^4q_1\cdots d^4q_{N_{\beta}},
%\eeq
%where $M_{\beta\alpha}$ is essentially the non-trivial part of $S_{\beta \alpha}$
%\beq
%S_{\beta \alpha}=-2\pi i \delta^{(4)}(\sum p_i-\sum q_i)M_{\beta \alpha}.
%\eeq
%Note that $T$ and $V$ are playing the role of IR regulator and at the end we should send them to infinity. 

%%%%%%%%%%%%%%%%%%%%%%%%%%%%
\subsection{$p_1p_2\rightarrow q_1q_2$ cross section in $\frac{\lambda}{4!}\phi^4$}
%%%%%%%%%%%%%%%%%%%%%%%%%%%%

As an example, we will find the cross section of $p_1p_2\rightarrow q_1q_2$ where $p_i^2=0$. Using \eqref{Sphi4} and the definition \eqref{def_tildeM}, to first order in $\lambda$
\beq
\widetilde M=\frac{\lambda}{(2\pi)^5}\sqrt{\widetilde W(q_1)\widetilde W(q_2)}.
\eeq
As a result, cross section is given by
\beq\label{eq13}
d\sigma=\frac{\lambda^2}{4\left(2\pi\right)^4|p_1\cdot p_2|}\widetilde W(q_1)\widetilde W(q_2)\delta^{(4)}(p_1+p_2-q_1-q_2)d^4q_1d^4q_2.
\eeq
Let us constraint the outgoing particles to be only on-shell $q_i^2=0$. In this case $\widetilde W$ functions in \eqref{eq13} pick up a delta function and one can check that \eqref{eq13} for outgoing on-shell particles results in the usual cross section of $\lambda \phi^4$ in LQFT. However, if we constraint (at least) one of the outgoing particles to be off-shell with a {\it fixed} mass, the cross section becomes zero. Cross section over outgoing off-shell particles is only non-zero when the integration over continuum mass is also performed. We see the significance of this in the next section when considering the scattering of off-shell particles.
Due to the contribution of off-shell states, the total cross section \eqref{eq13} is increased compared to the local theory.

%%%%%%%%%%%%%%%%%%%%%%%%%%%%%%%%%%%%
%%%%%%%%%%%%%%%%%%%%%%%%%%%%%%%%%%%%

\subsection{Off-shell particles and cross section}\label{cross section}
%%%%%%%%%%%%%%%%%%%%%%%%%%%%%%%%%%%%
%%%%%%%%%%%%%%%%%%%%%%%%%%%%%%%%%%%%

In order to calculate the cross section of any scattering involving incoming off-shell particles, we make use of the fact that off-shell particles can be thought as a continuum of massive particles. 

This can be done by expressing the two-point function as a sum over massive two point functions:
\beq\label{B1}
W(x,y)=\int_0^{\infty} d\mu^2 \rho(\mu^2)\int \frac{d^4p}{(2\pi)^4}2\pi \theta(p^0)\delta(p^2+\mu^2) e^{ip\cdot(x-y)},
\eeq
where $\rho(-p^2)=\frac{\widetilde W(p)}{2\pi}$ for $p^0>0$. Note from \eqref{field_operator} that $\rho(\mu^2)=\delta (\mu^2)+\widetilde \rho(\mu^2)$ where $\widetilde \rho$ is a finite function. In other words,
\bea
&&W(x,y)=\int \frac{d^4p}{(2\pi)^4}2\pi \theta(p^0)\delta(p^2) e^{ip\cdot(x-y)}\notag\\
&&+\int_0^{\infty} d\mu^2 \widetilde \rho(\mu^2)\int \frac{d^4p}{(2\pi)^4}2\pi \theta(p^0)\delta(p^2+\mu^2) e^{ip\cdot(x-y)}.~~~~\label{B2}
\eea
In order to make everything more similar to LQFT, we discretize the mass parameter to get
\bea
&&W(x,y)=\int \frac{d^4p}{(2\pi)^4}2\pi \theta(p^0)\delta(p^2) e^{ip\cdot(x-y)}\notag\\
&&+ \sum_{j=1}^{\infty} \Delta \mu^2 \widetilde \rho(\mu_j^2)\int \frac{d^4p}{(2\pi)^4}2\pi \theta(p^0)\delta(p^2+\mu_j^2) e^{ip\cdot(x-y)},~~~~\label{B3}
\eea
where $\mu_j^2 =j \Delta \mu^2$. \eqref{B3} is the same as \eqref{B2} in the limit $\Delta \mu^2\rightarrow 0$. 

The following field operator will yield the above two point function
\bea
&&\widehat \phi(x)= \int \frac{d^3\bold p}{(2\pi)^{3/2}}\frac{1}{\sqrt{2|\bold p|}}\left.\left(\widehat a _{\bold p,0}e^{ip\cdot x}+c.c\right) \right|_{p^0=|\bold p|}\notag\\
&&+ \sum_{j=1}^{\infty}\sqrt{\Delta\mu^2 \widetilde \rho(\mu_j^2)}  \int \frac{d^3\bold p}{(2\pi)^{3/2}\sqrt{2E_{\bold p ,\mu_j}}}  \left(\widehat a _{\bold p,\mu_j}e^{ip\cdot x}+c.c\right)\notag
\eea 

where
\bea
&E_{\bold p, \mu}=\sqrt{\bold p^2 +\mu^2}\\
&\left[\widehat a_{\bold p, \mu_i},\widehat a^{\dagger}_{\bold q,\mu_j}\right]=\delta^{(3)}(\bold p-\bold q)\delta_{\mu_i,\mu_j}\\
&\widehat a_{\bold p,\mu}|0\rangle=0
\eea
and state $|\bold p,\mu\rangle\equiv \widehat a^{\dagger}_{\bold p, \mu}|0\rangle$ is a one particle state with momentum $\bold p$, mass $\mu$ and energy $E_{\bold p,\mu}$. 

From now on, consider a concrete example of 2-2 scattering with $\frac{\lambda}{4!}\widehat \phi^4$ interaction and incoming particles with definite mass and momentum. The idea behind this proof can be generalized to more complicated examples. Up to first order in $\lambda$
\bea
\langle&& \mathbf {p_1},m_1;\mathbf{p_2},m_2|\widehat S|\mathbf{ q_1},\mu_1;\mathbf{ q_2},\mu_2\rangle=\notag\\
&&-\frac{i\lambda}{(2\pi)^2}\delta^{(4)}\left(\sum p-\sum q\right)\sqrt{\prod_{i=1}^2 \frac{(\Delta \mu^2)^2\rho(\mu_i^2)\rho(m_i^2)}{4E_{\mathbf{ q_i},\mu_i}E_{\mathbf{p_i},m_i}}}.~~~~~\label{S1}
\eea
In \eqref{S1}, if any of the particles was on-shell (say $\mu_1=0$), we should set $\Delta \mu^2 \rho(\mu_1^2)=1$, otherwise $\rho$ is replaced by $\widetilde \rho$.

The differential cross section is given by
\bea\label{B4}
d\sigma=(2\pi)^{-2}&&\lambda^2\frac{(\Delta \mu^2)^4\rho(\mu_1^2)\rho(\mu_2^2)\rho(m_1^2)\rho(m_2^2)}{16 u E_{\mathbf{p_1},m_1}E_{\mathbf{ p_2},m_2}E_{\mathbf{ q_1},\mu_1}E_{\mathbf{ q_2},\mu_2}}\notag\\
&&\delta^{(4)}(p_1+p_2-q_1-q_2)d^3\mathbf{p_1}d^3\mathbf{p_2}.
\eea
In order to get the total cross section, we should also sum over the mass parameter in the phase space of outgoing particles. In the (mass) continuum limit this means
\beq 
\sum\Delta\mu^2 \rho(m_i^2)\rightarrow\int dm_i^2\rho(m_i^2)
\eeq
which absorbs two factor of $\Delta \mu^2$ in \eqref{B4}; however, there are two remaining factors of $\Delta \mu^2$. If the incoming particles (even one of them) are off-shell, since $\rho(\mu^2)$ is a finite number, in the limit $\Delta \mu^2\rightarrow 0$, the cross section becomes zero. This means that the (total) transition rate of scattering with off-shell particles with {\it fixed} mass is zero. The cross section is only non-zero when both of the incoming particles are on-shell.

This is, in fact, consistent with what we have found in the previous section. There, we have shown that the transition rate of on-shell $\to$ off-shell is non-zero, only when the integration over mass of the off-shell particles is performed. In fact, scattering transition rate of on-shell particles to off-shell particles with fixed masses is zero. Since the theory is time reversal invariant, this suggests that the scattering transition rate of off-shell particles with fixed masses must be zero too; consistent with what we have found here. 

This also means that an initial state with a suitable continuum superposition of off-shell masses can scatter into on-shell modes (time reverse of the process of on-shell scattering into off-shell). However, as we argue in the next Section, these states are fine-tuned and generally we do not expect to find the system in these superpositions.

\subsection{Off-shell $\to$ on-shell scattering: continued} \label{offon_scatter}

In the previous section, we showed that the transition rate of scattering with off-shell particle(s) is zero. However, a suitable continuum superposition of off-shell particles can scatter non-trivially. In this section, we want to explain this point to a greater extent and argue that it is unlikely to find the system in these superpositions. We will not go through the detail of calculations since it is not essential to our argument in this secion.

We make use of the following toy model theory that mimics many properties of the proposed nonlocal theory:
\bea
\mathcal{L}&=&\frac{1}{2} \psi_0\Box \psi_0+\sum_{i=1}^{N}\frac{1}{2}\psi_{m_i}(\Box-m_i^2)\psi_{m_i}-\lambda\psi^4,\label{toymodel1}\\
\psi&\equiv&\psi_0+\sum_{i=1}^{N}\frac{g_i}{\sqrt{N}}\psi_{m_i}\notag.
\eea 
This is a theory of one massless scalar field (playing the role of on-shell modes) in addition to $N$ massive scalar fields (playing the role of off-shell modes) and we are interested in $N\to \infty$ limit of the theory ($\lambda$ and $g_i$'s are coupling constants and do not scale with $N$). The advantage of working with this theory is that while its behaviour is very similar to the non-local theory, \eqref{toymodel1} is a local quantum field theory and possibly more comprehensible to the reader.
The interaction term in \eqref{toymodel1} is designed in a way that interactions with massive (off-shell) fields are suppressed by a factor of $\sqrt{N}$ and in $N\to \infty$ limit their interactions become negligible. On the other hand, the number of off-shell fields goes to infinity. In what follows, we explain that this theory imitate many properties of off-shell and on-shell particles in the non-local theory.

First, let us define the following quantities: $\sigma^{\vec p_1\vec p_2}_{m_1m_2\to \mu_1\mu_2}$ is the scattering cross section of two particles with masses and momenta $m_1,\vec p_1$ and $m_2, \vec p_2$ into two particles with masses $\mu_1$ and $\mu_2$ ($\psi_{\mu_1}$ and $\psi_{\mu_2}$) and $\sigma^{\vec p_1\vec p_2}_{m_1m_2}$ is the total scattering cross section of two particles with masses and momenta $m_1,\vec p_1$ and $m_2, \vec p_2$.

Consider the scattering of two $\psi_0$ particles into two final particles. If we restrict the two final particles to be massive (off-shell fields with {\it fixed} masses), then the scattering cross section in $N\to \infty$ limit goes to zero. However, if we sum over all massive final states (all off-shell particles), the total cross section is non-zero. In fact, for different final states the corresponding cross sections scales with $N$ as follows:
\bea
\sigma^{\vec p_1\vec p_2}_{00\to00}~&&\propto N^0,\notag\\
\sigma^{\vec p_1\vec p_2}_{00\to0m}~&&\propto \frac{1}{N},~~m\neq 0,\notag\\
\sigma^{\vec p_1\vec p_2}_{00\to m_1m_2}~&&\propto \frac{1}{N^2},~~ m_1,m_2\neq 0.\notag
\eea
While the interactions with individual massive fields are suppressed, the number of massive states scales with $N$. In this way, the {\it total} scattering cross section of two initial massless particles into two massive final states, summed over all masses, is finite and non-zero (the same scaling works for scattering into one massless and one massive particles). 

On the other hand, any scattering with (at least) one massive initial state result into zero cross section. For example, the following total scattering cross sections (summed over all final states) scale with $N$ as  
\bea
\sigma^{\vec p_1\vec p_2}_{0m}~&&\propto \frac{1}{N},~~m\neq 0,\label{sigma0m}\\
\sigma^{\vec p_1\vec p_2}_{m_1m_2}~&&\propto \frac{1}{N^2},~~m_1,m_2\neq 0,
\eea
 and they vanish in $N\to \infty$ limit.
 
 As we showed, massive particles in this theory \eqref{toymodel1} mimic the properties of off-shell states in the non-local theory; they can be produced by the scattering of massless states, while the reverse process (scattering of massive states into massless) does not happen. 
 
 However, the theory is (obviously) time reversal invariant and massive $\to$ massless scatterings must take place. This is indeed true, but as we demonstrate here the initial massive state that scatters non-trivially must be a superposition of different masses. Consider state $\gamma$, a superposition of $M$ different masses, scatters off a massless particle. Then, the total transition probability $\Gamma_{0\gamma}$ scales as
\beq
\Gamma_{0\gamma}< A \frac{M}{N}
\eeq
Where $A$ have no dependence on $M$ and $N$ (see Appendix \ref{quantum_transition} for proof). This transition probability is non-zero in $N\to \infty$ limit, only when $M$ also scales with $N$. 

So, massive $\to$ massless scattering indeed happens. However, the massive state that scatters non-trivially must be a superposition of (infinitely) many different masses and in this sense is fine-tuned. It is similar to an egg that smashes into pieces upon falling on the ground; the reverse process of pieces assembling an egg can in principle happen, but it is very unlikely. 

In this sense, we expect the off-shell to on-shell scattering in the non-local theory to be negligible. In principle this transition can happen, but it is very implausible.
The essence of our reasoning in this section is based on thermodynamical arguments and although it is not a complete proof, we hope that we have provided enough evidence to show that off-shell $\to$ on-shell scattering is very unlikely. Definitely, further quantitative studies are needed to augment (or disprove) our claim. Perhaps, a good starting point is to consider the toy model theory \eqref{toymodel1}, since it shares a lot of properties of the non-local theory.

%%%%%%%%%%%%%%%%%%%%%%%%%%%%%%%%%%%%%%%
%%%%%%%%%%%%%%%%%%%%%%%%%%%%%%%%%%%%%%%
%%%%%%%%%%%%%%Massive Fields%%%%%%%%%%%%%%%%%%%%%
%%%%%%%%%%%%%%%%%%%%%%%%%%%%%%%%%%%%%%%
%%%%%%%%%%%%%%%%%%%%%%%%%%%%%%%%%%%%%%%

%%%%%%%%%%%%%%%%%%%%%%%%%%%%%%%%%%%%%%%
\section{Extension to Massive Scalar Fields}
\label{massive}
%%%%%%%%%%%%%%%%%%%%%%%%%%%%%%%%%%%%%%%

Throughout the paper, we only considered the modification of a massless scalar field. But what about massive scalar fields? One may suggest to replace $\Box$ with $\widetilde \Box$ in the equation of motion of a massive scalar field as follows
\beq\label{EOMmassive}
(\widetilde \Box-M^2)\phi(x)=J(x)
\eeq
and follow similar steps of quantization. However, this method does not work. If $M$ is a real number, then there is no mode satisfying \eqref{EOMmassive} in the absence of $J$. In other words, there is no on-shell modes.% and we don't recover the local EOM at low energies. 

Another way is to choose $M$ to be a complex number such that for a time-like future directed momentum $p$, $B(p)=M^2$. In this case, the mass of on-shell mode is given by $m^2\equiv -p^2$. However, $\widetilde \Box -M^2$ is no longer a real operator and the solution to \eqref{EOMmassive} generically cannot be real.

The extension to massive scalar fields can be done by considering the following observation. All of the properties in massless case can be read from the analytic structure of $B(p)$ in Figure \ref{analytic}. Massless modes are on-shell because there is a simple zero at $p^2=0$ and there are off-shell modes for time-like momenta because there is a cut for time-like momenta in \ref{analytic}.

 In this way, the extension to massive case seems much simpler. $\Box -m^2$ must be replaced with $\widetilde \Box_m$ whose eigenvalues $B_m(p)$ satisfy the followings:
\begin{enumerate}
\item There is only one simple zero at $p^2=-m^2$. Also $\lim_{p^2+m^2 \to 0} \frac{B_m(p)}{p^2+m^2}=-1$ to get the correct local limit.
\item The cut must be only on momenta with higher masses $p^2<-m^2$. Otherwise, in the quantum theory, there are off-shell modes with mass smaller than $m$ which makes the on-shell mode unstable (on-shell modes can always decay into off-shell modes with less mass).
\item ${\rm Im} B_m(p)\ge 0$ for $p^0>0$. 
\end{enumerate}
Conditions 4 and 5 in Section \ref{mBoxDef} and \eqref{sgn} must be replaced by the above-mentioned conditions. One easy way to come up with such an operator is to make use of the existing operator $B(p)$ in the massless case, and consider it as a function of $p^2$ and $sgn(p^0)$. Then,
\beq
B_m(p)=B(p^2+m^2,sgn(p^0)).
\eeq
has all the desired properties (this also has been shown in \cite{Belenchia:2014fda}).

%%%%%%%%%%%%%%%%%%%%%%%%%%%%%%%%%%%%%%%
%%%%%%%%%%%%%%%%%%%%%%%%%%%%%%%%%%%%%%%
%%%%%%%%%%%%%%Conclusion%%%%%%%%%%%%%%%%%%%%%
%%%%%%%%%%%%%%%%%%%%%%%%%%%%%%%%%%%%%%%
%%%%%%%%%%%%%%%%%%%%%%%%%%%%%%%%%%%%%%%

%%%%%%%%%%%%%%%%%%%%%%%%%%%%%%%
\section{Conclusion}
\label{conclusion}
%%%%%%%%%%%%%%%%%%%%%%%%%%%%%%%
In this paper, we studied the physical consequences of a causal non-local evolution of a massless scalar field. We started by modifying the d'Alembertian to a causal non-local operator at high energies.
Quantization of a free field showed that the field represents a continuum of massive particles. In fact, there were two sets of modes: on-shell modes (massless particles) and off-shell modes (massive particles). 

The Feynman rules for the perturbative calculation of S-matrix amplitudes were discussed.  The most important result (in our opinion) is the fact that the cross section of any scattering with off-shell particles is zero. This suggests that although these modes exist and probably can be detected by other means, there is no way of detecting them through scattering experiments. This property opens up the possibility that dark matter particles might be just the off-shell modes of known matter. Finally, we extended this formalism to massive scalar fields.

Throughout this paper we only considered scalar field theories, but how about other types of fields?
Extension to other types of fields, such as vector field, is not as straightforward as for scalar fields. Incorporating gauge symmetry in the theory is another important issue. Whether causal Lorentzian evolution can be extended to vector fields (and other types fields rather than scalars) can be the subject of future studies.

%%%%%%%%%%%%%%%%%%%%%%%%%%%%%%%%%%%%%
\acknowledgments
We thank Rafael Sorkin, Niayesh Afshordi, Dionigi Benincasa and Gregory Gabadadze for useful discussions throughout the course of this project.

%This research was supported in part by NSERC through grant RGPIN-418709-2012.
%
This research was supported in part by Perimeter Institute for
Theoretical Physics. Research at Perimeter Institute is supported by the
Government of Canada through Industry Canada and by the Province of
Ontario through the Ministry of Research and Innovation.

\bibliography{Causal_Lorentzian_Evolution}

%%%%%%%%%%%%%%%%%%%%%%%%%%%%%%%%%%%%
%%%%%%%%%%%%%%%%%%%%%%%%%%%%%%%%%%%%
%%%%%%%%%%%%%%%%%%%%%%%%%%%%%%%%%%%%
\appendix

%==========================================================
\section{Existence and Examples of $\widetilde \Box$}
\label{ret_Box}
%==========================================================
Here we will show there are operators $\widetilde \Box$ which satisfy all the axioms
introduced in Section \ref{mBoxDef}. In fact, we will outline a
procedure for constructing such operators. 

We shall consider the following operator:
\beq
	\Lambda^{-2}(\widetilde \Box\phi)(x)=a\phi(x)+\Lambda^4\int_{J^{-}(x)}f(\Lambda^2\tau_{xy}^2)\phi(y)d^4y,
\eeq
where $\Lambda$ denotes the nonlocality energy scale,
$a$ is a dimension-less real number, $J^-(x)$ denotes the causal past of $x$, and
$\tau_{xy}$ is the Lorentzian distance between $x$ and $y$: 
\beq
	\tau_{xy}^2=(x^0-y^0)^2-|\vecx-\vecy|^2.
\eeq
%\beq
%\tau_{xy}^2=(x^0-y^0)^2-|\vecx-\vecy|^2, \qquad 
%y=(y^0,\vecy), \qquad
%x=(x^0,\vecx).
%\eeq
It may be shown that
\begin{align}
	\widetilde \Box e^{ip\cdot x}&=B(p)e^{ip\cdot x},\\
	B(p)&=\Lambda^{2}\widetilde{g}(p/\Lambda),\\
	\widetilde{g}(z)&=a+\int_{J^{+}(0)}f((y^0)^2-|\vecy|^2)e^{-iz\cdot y}d^4y,
\end{align}
where as usual $x\cdot y=\eta_{\mu\nu}x^{\mu}y^{\nu}$. 
Evaluating $\widetilde{g}(z)$ amounts to computing the Laplace transform of a retarded, Lorentz
invariant function, which has been done in \cite{DLMF}. It follows from their result that
\begin{align}
	\widetilde{g}(z)&=g(z\cdot z),\label{ret_Box1}\\
	g(Z)&=a+4\pi Z^{-\frac{1}{2}}\int_{0}^{\infty}f(s^2)s^{2}K_{1}(Z^{1/2}s)ds\label{integralgZ},
\end{align}
where an infinitesimal time-like and future-directed imaginary part ought to be added 
to $z$ on the right hand side of \eqref{ret_Box1} (see \cite{Aslanbeigi:2014zva} for more details).

%--------------------------------------------------------------------------------------------------------
\subsection{IR conditions}
%--------------------------------------------------------------------------------------------------------
The infrared condition \eqref{IRcondition} is equivalent to satisfying
\beq\label{IRgZ}
	g(Z)\xrightarrow{Z\rightarrow 0}-Z.
\eeq
In \cite{Aslanbeigi:2014zva}, a framework is developed to determine
what constraints \eqref{IRgZ} places on $a$ and $f$, for some specific
choices of $f$ which arise in causal set theory.
Generalizing that methodology in a straightforward manner, we 
find that \eqref{IRgZ} is true if and only if the following conditions are satisfied:
\begin{align}
	\int_{0}^{\infty}f(s^2)s^{2k+1}ds&=0, \qquad k=0,1,2\label{IRmain} \\
	\int_{0}^{\infty}f(s^2)s^{5}\ln s ds&=-\frac{4}{\pi},\label{IRa} \\
	a+2\pi\int_{0}^{\infty}f(s^2)s^{3}\ln s ds&=0.\label{IRnorm}
\end{align}

%--------------------------------------------------------------------------------------------------------
\subsection{From $B(p)$ to $\widetilde \Box$}
\label{BtoBox}
%--------------------------------------------------------------------------------------------------------
It is often desirable to constrain the behaviour of $B(p)$, as opposed to $\widetilde \Box$ directly.
For instance, as is argued in Section \ref{Hilbert_representation}, the quantum theory is well behaved
only when the imaginary part of $B(p)$ (for timelike and future-directed $p$)
is always positive. The question then becomes: {\it are there any choices of $a$ and 
$f$ which allow for this possibility, provided the IR conditions \eqref{IRmain}--\eqref{IRnorm}
are satisfied}? 
To answer this question, we turn the problem around.
Given a choice of $B(p)$, we reconstruct $a$ and $f$ and then ask if the IR conditions are met.

It can be shown that for $x>0$: (see e.g. $10.27.9$ and $10.27.10$ of \cite{DLMF})
\begin{align}
g(-x^2-i\epsilon)&=g_R(-x^2-i\epsilon)+ig_I(-x^2-i\epsilon),\\
g_R(-x^2-i\epsilon)&=a+\frac{2\pi}{x}\int_{0}^{\infty}f(s^2)s^2Y_1(xs)ds,\\
g_I(-x^2-i\epsilon)&=-\frac{2\pi^2}{x}\int_{0}^{\infty}f(s^2)s^2J_1(xs)ds.
\label{ftogi}
\end{align}
We can now use the following orthonormality conditions of Bessel functions (see e.g. $1.17.13$ of 
\cite{DLMF}) to express $f$ in terms of $\widetilde{g}_I$:
\beq
\delta(x-\widetilde{x})=x\int_{0}^{\infty}tJ_1(xt)J_1(\widetilde{x}t)dt.
\eeq
Doing so yields:
\begin{align}
f(s^2)&=f_g(s^2)+h(s^2),\\
f_g(s^2)&=-\frac{1}{2\pi^2s}\int_{0}^{\infty}g_I(-x^2-i\epsilon)x^2J_1(sx)dx,
\label{gtof}
\end{align}
where $h$ satisfies for all x:
\beq
\int_{0}^{\infty}h(s^2)s^2J_1(xs)ds=0.
\label{hcond}
\eeq
This means that specifying $\widetilde{g}_I(-x^2-i\epsilon)$ fixes $f$ up to any part for which the 
right hand side of \eqref{ftogi} vanishes. 
One example of a nontrivial function which satisfies \eqref{hcond}
is the delta function: $h(x)=\delta^+(x)\equiv\delta(x-\epsilon)$, where 
$\epsilon$ is an arbitrarily small positive real number.

We can now express the IR conditions in terms of $g_I$ and $h$:
\bea
&&\int_{0}^{\infty}h(s^2)s^{2k+1}ds\notag\\
&&-\frac{1}{2\pi^2}\int_{0}^{\infty}g_I(-x^2-i\epsilon)x^2\int_{0}^{\infty}ds s^{2k}J_1(xs)=0,\\
&&\int_{0}^{\infty}h(s^2)s^5\ln sds\notag\\
&&-\frac{1}{2\pi^2}\int_{0}^{\infty}g_I(-x^2-i\epsilon)x^2\int_{0}^{\infty}ds s^4J_1(xs)\ln s=-\frac{4}{\pi},~~~~~~~\\
a&&+2\pi\int_{0}^{\infty}h(s^2)s^3\ln sds\notag\\
&&-\frac{1}{\pi}\int_{0}^{\infty}g_I(-x^2-i\epsilon)x^2\int_{0}^{\infty}ds s^2J_1(xs)\ln s=0.
\eea
The above integrals over $s$ are not absolutely convergent, so use the usual trick:
\begin{align}
	\int_{0}^{\infty}ds J_1(xs)e^{-\delta s}&\xrightarrow{\delta\to0}\frac{1}{x},\\
	\int_{0}^{\infty}ds s^2J_1(xs)e^{-\delta s}&\xrightarrow{\delta\to0}\frac{3\delta}{x^4},\\
	\int_{0}^{\infty}ds s^4J_1(xs)e^{-\delta s}&\xrightarrow{\delta\to0}\frac{-45\delta}{x^6},\\
	\int_{0}^{\infty}ds s^{2}J_1(xs)\ln s e^{-\delta s}&\xrightarrow{\delta\to0}-2x^{-3},\\
	\int_{0}^{\infty}ds s^{4}J_1(xs)\ln s e^{-\delta s}&\xrightarrow{\delta\to0}16x^{-5}.
\end{align}
Having the delta function example in mind, we shall require $h$ to satisfy for all $k=1,2$
\beq
	\int_{0}^{\infty}h(s^2)s^{2k+1}ds=0,\qquad
	\int_{0}^{\infty}h(s^2)s^{2k+1}\ln s ds=0,
\eeq
Also, we assume that the following integrals converge:
\begin{align}
	\left|\int_{0}^{\infty}g_I(-x^2-i\epsilon)x^{-k}dx\right|&<\infty, \qquad k=1,2,3,4\label{convg1}\\
	\left|\int_{0}^{\infty}g_I(-x^2-i\epsilon)x^{-k}\ln xdx\right|&<\infty \qquad k=2,4.\label{convg2}
\end{align}
The IR conditions then reduce to
\begin{align}
	\int_{0}^{\infty}g_I(-x^2-i\epsilon)x~dx&=\pi^2\int_{0}^{\infty}h(u)du,\label{gtoIR1}\\
	\int_{0}^{\infty}g_I(-x^2-i\epsilon)x^{-3}dx&=\frac{\pi}{2},\label{gtoIR2}\\
	\int_{0}^{\infty}g_I(-x^2-i\epsilon)x^{-1}dx&=-\frac{\pi}{2}a.\label{gtoIR3}
\end{align}
Note that the only nontrivial condition to satisfy is \eqref{gtoIR1}, since \eqref{gtoIR2} just fixes the normalization of 
$g_I$ and \eqref{gtoIR3} determines $a$. Note that for positive $g_I(-x^2-i\epsilon)$ which is required by consistent quantum theory, $a$ must be a negative number.

If $h$ is taken to be zero, then $g_I$ ought to change sign, which leads to a quantum theory with an unbounded Hamiltonian.
We note that the class of operators which arise in causal set theory in \cite{Aslanbeigi:2014zva} all have $h=0$, and therefore this feature.

Let us work out a complete example in 4D. Let
\beq
g_I(-x-i\epsilon)=Ax^2e^{-x/2}, \qquad
h(x)=\alpha\delta^+(x).
\eeq
where $A$ and $\alpha$ are real constants. 
It can then be shown using \eqref{gtoIR1}--\eqref{gtoIR3}:
\beq
A=\frac{\pi}{2}, \qquad \alpha=\frac{4}{\pi}, \qquad a=-2.
\eeq
It then follows from \eqref{gtof} that
\beq
f_g(s)=-\frac{e^{-s/2}}{4\pi}(24-12s+s^2).
\eeq
Therefore:
\beq
f(s)=\frac{4}{\pi}\delta^+(s)-\frac{e^{-s/2}}{4\pi}(24-12s+s^2).
\eeq

%--------------------------------------------------------------------------------------------------------
\subsection{Stability from positivity of $g_I$}
%--------------------------------------------------------------------------------------------------------
We have required that evolution defined by $\widetilde \Box$ should be stable.
Instabilities
are in general associated with ``unstable modes'', 
and in line with \cite{Aslanbeigi:2014zva}, we shall use
this as our criterion of instability.
More specifically, we take such a mode to be a plane-wave
$e^{ip\cdot x}$ satisfying the equation of motion
$\widetilde \Box e^{ip\cdot x}=0$, with the wave-vector $p$ possessing
a future-directed timelike imaginary part (i.e. $p=p_R+ip_I$ where
$p_I\cdot p_I<0$ and $p_I^0>0$). It is shown in \cite{Aslanbeigi:2014zva}
that the necessary and sufficient condition for avoiding unstable modes is
\beq
	g(Z)\neq0  \ ,\qquad \forall\ Z\neq0 \text{ and } Z\in\mathbb{C}.
\eeq

On the other hand, we argued in \ref{Hilbert_representation} that for consistency reasons we need to assume $ImB(p)>0$ for $p^0>0$ which implies $g(Z)$ has a positive (negative) imaginary part under (above) the cut in Figure \ref{analytic}. 

Here, we show that not even stability condition and positivity of $g_I(-x^2-i\epsilon)$ (see Appendix \ref{BtoBox}) are consistent, but latter is a sufficient condition for stability.  In order to prove it, we make the following assumptions:
\begin{enumerate}
\item $g(Z)$ has a simple zero at $Z=0$. IR conditions on $g(Z)$ \eqref{IRgZ} guarantee this assumption.
\item  $g(Z)$ has positive (negative) imaginary part under (above) the cut.
\end{enumerate}

\begin{figure}
        \centering
        	\includegraphics[width=0.8\hsize]{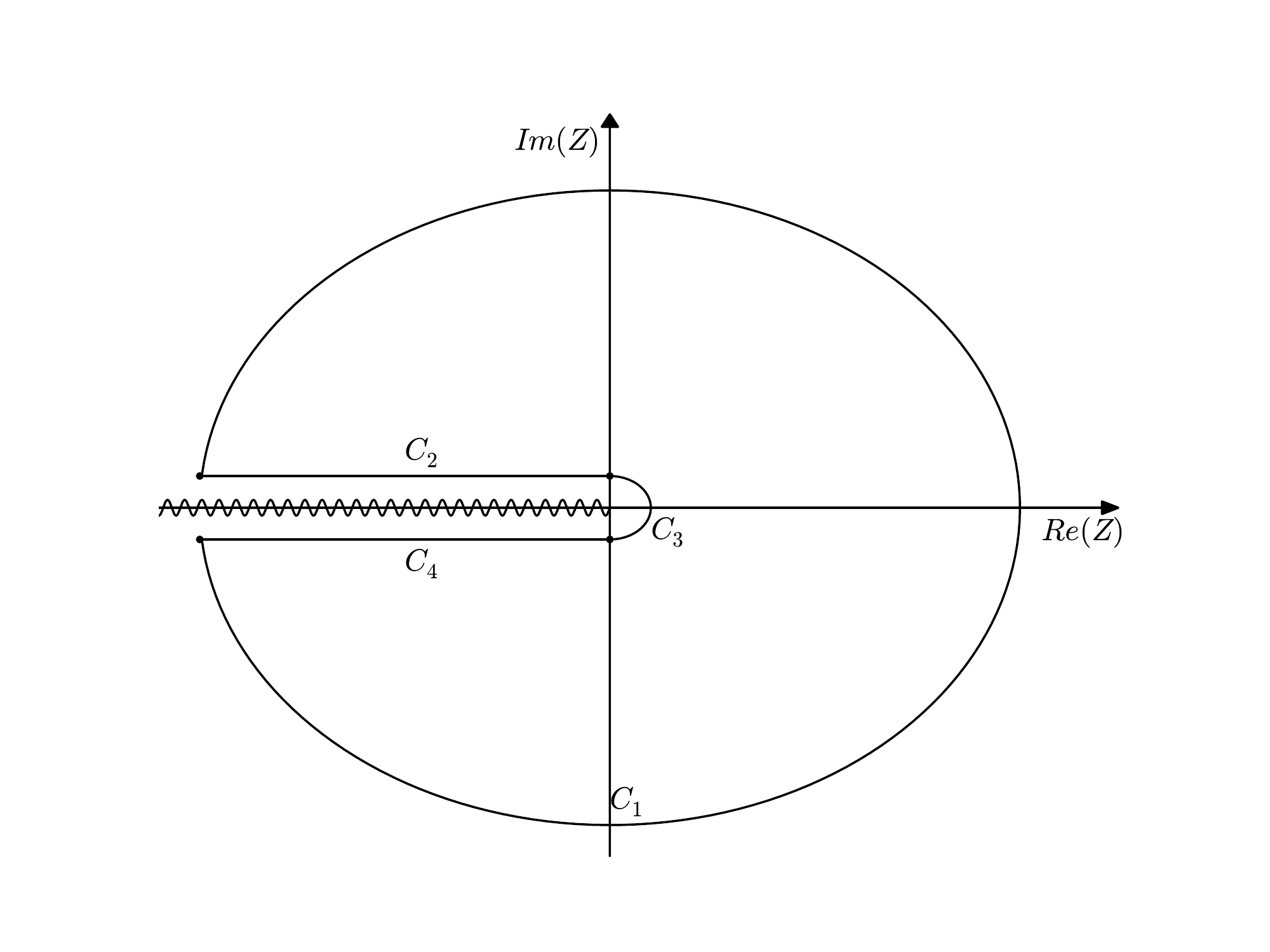}
	 \caption{The integration path in the complex $Z$ plane. The closed contour is taken to be counterclockwise.}
	\label{stabcontour}
\end{figure}

We prove this by counting the number of zeros of $g$ inside contour $C=C_1+C_2+C_3+C_4$ in Figure \ref{stabcontour}. 

If $N$ and $P$ are the number of zeros and poles of $g$, respectively, inside the contour $C$ (taken to be anticlockwise), then
\beq
\label{Zeros}
\int_CdZ~\frac{g'(Z)}{g(Z)}=-2\pi i (N-P).
\eeq
Let's evaluate the left hand side of \eqref{Zeros} for each contour separately:
\begin{enumerate}

\item $C_1$: According to \eqref{integralgZ}, $g(Z)$ approaches the constant value of $a<0$ (see \ref{BtoBox}) for large $Z$. In fact, $g(Z)\to a+\mathcal{O}(\frac{1}{Z^n})$ for some positive value of $n$ (which depends on the function $f$). This means for $a\neq 0$, 
\beq
\int_{C_1}dZ~\frac{g'(Z)}{g(Z)}=0.
\eeq

\item $C_2$ \& $C_4$: Since the values of $g$ above and under the cut are complex conjugate of each other, the contribution from these diagrams can be added together to get
\bea
\label{C24contours}
\int_{C_2+C_4}dZ~\frac{g'(Z)}{g(Z)}=&&2i{\rm Im}\int_{-\infty}^0dx~\frac{g'(x+i\epsilon)}{g(x+i\epsilon)}\notag\\
=&&2i{\rm Im}\ln\left[\frac{g(0+i\epsilon)}{g(-\infty+i\epsilon)}\right],
\eea
where $\epsilon$ is an infinitesimal positive number.

If we define $g(Z)=r_g(Z)e^{i\varphi_g(Z)}$, the right hand side of \eqref{C24contours} (apart from the factor of $2i$) measures how much $\varphi_g$ rotates from $Z=-\infty+i\epsilon$ to $Z=0+i\epsilon$. Since ${\rm Im}g(x+i\epsilon)<0$ on the whole negative real line, $\ln\left[g(x+i\epsilon)\right]$ is definable on one Riemann sheet. Combining this result with $g(-\infty+i\epsilon)=a<0$ and $g(0+i\epsilon)=-i\epsilon$, we get
\beq
\int_{C_2+C_4}dZ~\frac{g'(Z)}{g(Z)}=i\pi.
\eeq 

\item $C_3$: IR conditions require that close to $Z=0$, $g(Z)=-Z$. This means 
\beq
 \int_{C_3}dZ~\frac{g'(Z)}{g(Z)}=\int_{C_3}~\frac{1}{Z}=-i\pi.
 \eeq

\end{enumerate}
Adding the values of all the contours and considering the fact that $g(Z)$ is finite everywhere ($P=0$), we conclude that the number of zeros of $g$ in complex plane of $Z$ (inside contour $C$) is zero. Since there is no zero on the negative real line (${\rm Im}g(x+i\epsilon)\neq 0$), there is no zero of $g$ in the complex plane of $Z$ except the one at $Z=0$. Therefore, stability has been proven.

%==========================================================
\section{FDT}
\label{AppFDT}
%==========================================================
Here, we present the proof of \eqref{FDT} \footnote{Most of the content of this appendix is taken from \cite{birrell1984quantum}.}. Let's start by the following definitions
\bea
i\Delta(x,y)&&\equiv \left[ \widehat \phi(x),\widehat \phi(y)\right],\\
G^{(1)}(x,y)&&\equiv \left \langle \left\{\widehat \phi(x),\widehat \phi(y)\right\}\right \rangle,\\
W^+(x,y)&&\equiv \left \langle \widehat \phi(x)\widehat \phi(y)\right \rangle,\\
W^-(x,y)&&\equiv \left \langle \widehat \phi(y)\widehat \phi(x)\right \rangle,\\
G^F(x,y)&&\equiv -i \left \langle T\widehat \phi(x)\widehat \phi(y)\right \rangle,
\eea
where $\{\}$ is anti-commutator and $\langle \rangle$ shows expectation value in a quantum state. If we define
\beq
G^R(x,y)\equiv \Delta(x,y) H(x\succ y),\\
\eeq
\beq
G^A(x,y)\equiv -\Delta(x,y) H(x \prec y),\footnote{where $H$ is the Heaviside function: $H(x\succ y)=1$ if $x \succ y$ and otherwise 0 }
\eeq
we get the following relations
\bea
i\Delta(x,y)&&=W^+(x,y)-W^-(x,y)\notag\\
&&=i\left[G^R(x,y)-G^A(x,y)\right]\label{app1},\\
G^{(1)}(x,y)&&=W^+(x,y)+W^-(x,y),\\
G^A(x,y)&&=G^R(y,x),\\
G^F(x,y)&&=\frac{1}{2}\left[G^R(x,y)+G^A(x,y)\right]-\frac{i}{2}G^{(1)}(x,y).~~~~~~~
\eea
For a translational invariant system, the value of all the two point functions depend only on space-time separation. This will allow us to define the following Fourier transform with respect to time
\beq
\xbar A(\omega,\bold x,\mathbf{x'})\equiv \int dt~ A(t,\bold x;t',\mathbf{x'})e^{-i\omega (t-t')}.
\eeq
Now, let us assume that the quantum system is in thermal state with temperature $T=\frac{1}{\beta}$. It requires that 
\beq
W^{\pm}(t,\bold x;t',\mathbf{x'})=W^{\mp}(t+i\beta,\bold x;t',\mathbf{x'}),
\eeq
resulting in the following relation in Fourier space
\beq
\xbar W^+(\omega,\bold x,\bold y)=e^{\beta \omega}\xbar W^-(\omega, \bold x, \bold y).
\eeq
Using \eqref{app1}, we get
\bea
\xbar W^+(\omega,\bold x,\bold y)=\frac{i\xbar \Delta (\omega,\bold x,\bold y)}{1-e^{-\beta \omega}},\\
\xbar W^-(\omega,\bold x,\bold y)=-\frac{i\xbar \Delta(\omega,\bold x,\bold y)}{1-e^{\beta \omega}}.
\eea
On the other hand, since $G^R$ and $G^A$ are time transpose of each other, in Fourier space they are complex conjugate. As a result,
\bea
{\rm Im} \xbar G^F(\omega, \bold x, \bold y)&=&-\frac{1}{2}{\rm Re}\xbar G^{(1)}(\omega, \bold x, \bold y)\notag\\
&=&-\frac{1}{2}\left[\xbar W^+(\omega,\bold x,\bold y)+\xbar W^-(\omega,\bold x,\bold y)\right]\notag\\
&=&-\frac{1}{2}i\xbar \Delta(\omega, \bold x, \bold y)\coth(\frac{\beta \omega}{2})\label{app3} 
\eea
where ${\rm Im}$ and ${\rm Re}$ are imaginary part and real part respectively and in the second line we have used the positivity of two point function $W^+$ (resulting that  $\xbar W^+(\omega,\bold x,\bold y)$ and $\xbar W^-(\omega,\bold x,\bold y)$ are real.)

With the assumption that this field theory in Hilbert space representation has an equivalent representation in terms of double path integral, time ordered two point function is given by \eqref{Tfunction}. In Fourier space, it reads
\beq
\xbar G^F(\omega, \bold x, \bold y)=\frac{1}{2}\left[\xbar G^K(\omega, \bold x, \bold y)+\xbar G^R(\omega, \bold x, \bold y)+\xbar G^A(\omega, \bold x, \bold y)\right].
\eeq
$\xbar G^K(\omega, \bold x, \bold y)$ is a total imaginary number and $\xbar G^R(\omega, \bold x, \bold y)+\xbar G^A(\omega, \bold x, \bold y)$ is a real number. As a result,
\beq\label{app4}
\xbar G^K(\omega, \bold x, \bold y)=2i {\rm Im} \xbar G^F(\omega, \bold x, \bold y).
\eeq
Combining \eqref{app1}\eqref{app3}\eqref{app4} we arrive at
\beq
\xbar G^K(\omega, \bold x, \bold y)=\coth(\frac{\beta \omega}{2})\left[\xbar G^R(\omega, \bold x, \bold y)-\xbar G^A(\omega, \bold x, \bold y)\right],
\eeq
which reduces to \eqref{FDT} at zero temperature.

\section{Quantum Transition}
\label{quantum_transition}
We start by proving a simple theorem for any quantum system. Consider a quantum mechanical system in the (normalized) initial state $|\alpha\rangle$ evolves in time and the probability of finding the system at a later time $t_f$ in the state $|\beta_i\rangle$ is called $P_i$, and assume $|\beta_i\rangle$'s are orthonormal:
\beq
P_i=|\langle\beta_i |U|\alpha\rangle|^2
\eeq
where $U$ is the time evolution operator.

Now, consider a (normalized) state $|\beta\rangle$ as a superposition of $|\beta_i\rangle$ states:
\bea
&&|\beta\rangle=\sum_i c_i |\beta_i\rangle\\
&&\sum_i |c_i|^2=1.\notag
\eea
Probability $P$ of measuring the system at time $t_f$ in the state $|\beta\rangle$ is given by
\beq
P=|\langle\beta |U|\alpha\rangle|^2.
\eeq
Then,
\bea
P&&=|\langle\beta |U|\alpha\rangle|^2=\left|\sum_ic^*_i\langle\beta_i|U|\alpha\rangle\right|^2\notag\\
&&\le \left(\sum_i|c_i|^2\right)\left(\sum_i|\langle\beta_i |U|\alpha\rangle|^2\right)\notag\\
&&=\sum_i P_i\label{boundP}
\eea
where we have used the triangular inequality in the second line. So $P$ is bounded from above by $\sum_i P_i$. 

Now, let's get back to the scattering of a massless particle with state $\gamma$, a superposition of $M$ different masses, in Section \ref{offon_scatter}. We already have shown (see \eqref{sigma0m}) that $\Gamma_{0m_i}$ defined as transition probability of a massless particle scattering with a massive particle (mass $m_i$) scales with $N$ as 
\beq
\Gamma_{0m_i}=\frac{A_i}{N}
\eeq
where $A_i$ depends on the momentum of the particles but independent of $N$. 
Using \eqref{boundP} for transition probabilities, we conclude that
\beq
\Gamma_{0\gamma}\le \sum_i\frac{A_i}{N}\le A\frac{M}{N}
\eeq
where $A$ is the maximum of $A_i$'s.

\end{document}